\documentclass[fleqn,usenatbib]{mnras}

\usepackage{newtxtext,newtxmath}
\usepackage{diagbox}
\usepackage{makecell}
\usepackage[justification=centering]{caption} 

\usepackage[T1]{fontenc}

\DeclareRobustCommand{\VAN}[3]{#2}
\let\VANthebibliography\thebibliography
\def\thebibliography{\DeclareRobustCommand{\VAN}[3]{##3}\VANthebibliography}

\usepackage{graphicx}	
\usepackage{amsmath}	
\usepackage{subcaption} 
\usepackage{multirow} 

\title[maximum mass of neutron stars]{What constraints can one pose on the maximum mass of neutron stars from multi-messenger observations?}

\author[Shunke Ai et al.]{
Shunke Ai$^{1}$\thanks{E-mail: shunke.ai@whu.edu.cn},
He Gao$^{2,3}$\thanks{E-mail: gaohe@bnu.edu.cn},
Yong Yuan$^{1}$,
Bing Zhang$^{4,5}$\thanks{E-mail: bing.zhang@unlv.edu}
and Lin Lan$^{6}$
\\
$^{1}$Department of Astronomy, School of Physics and Technology, Wuhan University, Wuhan 430072, China\\
$^{2}$Institute for Frontier in Astronomy and Astrophysics, Beijing Normal University, Beijing 102206, China\\
$^{3}$Department of Astronomy, Beijing Normal University, Beijing 100875, People's Republic of China\\
$^{4}$Nevada Center for Astrophysics, University of Nevada Las Vegas, Las Vegas, NV 89154, USA\\
$^{5}$Department of Physics and Astronomy, University of Nevada Las Vegas, Las Vegas, NV 89154, USA\\
$^{6}$CAS Key Laboratory of Space Astronomy and Technology, National Astronomical Observatories, Chinese Academy of Sciences, Beĳing 100101, China\\
}

\date{Accepted XXX. Received YYY; in original form ZZZ}

\pubyear{2015}

\begin{document}
\label{firstpage}
\pagerange{\pageref{firstpage}--\pageref{lastpage}}
\maketitle

\begin{abstract}
The maximum mass of neutron stars ($M_{\rm TOV}$) plays a crucial role in understanding their equation of state (EoS). Previous studies have used the measurements for the compactness of massive pulsars and the tidal deformability of neutron stars in binary neutron star (BNS) mergers to constrain the EoS and thus the $M_{\rm TOV}$. The discovery of the most massive pulsar, PSR J0952-0607, with a mass $\sim 2.35M_{\odot}$, has provided a valuable lower limit for $M_{\rm TOV}$. Another efficient method to constrain $M_{\rm TOV}$ is by examining the type of central remnant formed after a BNS merger. Gravitational wave (GW) data can provide the total mass of the system, while accompanying electromagnetic signals can help infer the remnant type. In this study, we combine all the previous constraints and utilize the observational facts that about $24\%$ of the short gamma-ray bursts are followed by an X-ray internal plateau, which indicate that roughly this fraction of BNS mergers yield supermassive neutron stars, to perform (Markov Chain) Monte Carlo simulations. These simulations allow us to explore the probability density distribution of $M_{\rm TOV}$ and other parameters related to BNS mergers. Our findings suggest that $M_{\rm TOV}$ is likely around $2.49M_{\odot} - 2.52M_{\odot}$, with an uncertainty range of approximately [$-0.16M_{\odot}$, $0.15M_{\odot}$] ([$-0.28M_{\odot}$, $0.26M_{\odot}$]) at $1\sigma$ ($2\sigma$) confidence level. Furthermore, we examine the type of merger remnants in specific events like GW170817 and GW190425 to further constrain $M_{\rm TOV}$ and other relevant parameters, which can help to understand the physical processes involved in BNS mergers. 
\end{abstract}




\section{Introduction}

The equation of state (EoS) for neutron stars (NSs) has long been an open question, although astronomical observations continues to tighten the allowed parameter space. Generally, constraints on the EoS for NSs come from two main perspectives: their compactness at a certain mass or radius, and their maximum mass under non-rotating conditions ($M_{\rm TOV}$). 

The compactness of a NS can be inferred through the jointly measuring its mass and radius. In the case of a known massive pulsar with precise mass measurement (usually done through Shapiro delay) \citep[e.g.][]{antoniadis2013,cromartie2020,romani2022}, its radius can be inferred from X-ray observations, such as those conducted by the X-ray observations by the Neutron star Interior Composition Explorer (NICER) \citep[e.g.][]{miller2019,riley2019,miller2021,riley2021}. Once the radius of a NS at a specific mass is constrained, any EoS predicting a radius outside the allowed region at this mass could be ruled out. The tidal deformability of NSs which reflects their compactness, can also be inferred from the GW waveform in the late-inspiral phase of binary neutron star (BNS) mergers \citep{flanagan2008,favata2014}. Constraints on the parameters related to the NS EoS has been derived from GW170817 \citep{abbott2017a,abbott2018} and GW190425 \citep{abbott2020}. \cite{Dietrich2020} made the first attempt to combine as many as multi-messenger observations, including the pulsar mass measurements of PSR J0740+6620, PSR J0348+4042, and PSR J1614–2230 \citep{Demorest2010,antoniadis2013,Arzoumanian2018,cromartie2020}, GW data from the NS mergers GW170817 and GW190425, information from the kilonova AT 2017gfo and
the gamma-ray burst GRB 170817A and its afterglow \citep{abbott2017a, abbott2017b, abbott2017c}, and the NICER observation of PSR J0030+0451 \citep{riley2019,miller2019}, to pose a joint constraint on the EoS of NSs. Under this framework, people continues to refine the constraints with updated observational data and models, e.g. adding the NICER and XMM-Newton measurements for PSR J0740+6620 \citep{riley2021,miller2021,Biswas2021,li2021,Annala2022,Somasundaram2022,Raaijmakers2021,Legred2021,Pang2021}. A comprehensive review on the previous related works is presented in \cite{Pang2021}. Generally, 
the existing constraints on the characteristic radius $R_{1.4}$ of NSs are much more stringent than those on $M_{\rm TOV}$, e.g. $R_{1.4} = 12.56^{+0.45}_{-0.40}{\rm km}$ and $M_{\rm TOV} = 2.27^{+0.34}_{-0.18}M_{\odot}$ with the uncertainties at $1\sigma$ confidence level in \cite{miller2021}.


A tight constraint on the maximum mass of NSs ($M_{\rm TOV}$) is sorely needed now, because along with the already well-constrained compactness, it becomes much easier to differentiate NS EoSs proposed in the literature. A direct lower limit on $M_{\rm TOV}$ is the mass of the most massive pulsar observed. For example, as mentioned above, several pulsars with masses greater the $2M_{\odot}$ have been confirmed, such as PSR J0348+0432 ($M = 2.01^{+0.04}_{-0.04}M_{\odot}$)  \citep{antoniadis2013}, PSR J0740+6620 ($M = 2.072^{+0.067}_{-0.066}M_{\odot}$) \citep{cromartie2020,riley2021} and PSR J0952-0607 ($M = 2.35^{+0.17}_{-0.17}M_{\odot}$)  \citep{romani2022}, with PSR J0952-0607 being the currently most massive one. Therefore, approximately $M_{\rm TOV} > 2.35M_{\odot}$ is obtained. 

Another method to constrain $M_{\rm TOV}$ is based on the type of central remnant of the BNS mergers. In theory, the type of merger remnant depends on $M_{\rm TOV}$ and the remnant mass $M_{\rm rem}$ \citep{rosswog2000,rezzolla2010,rezzolla2011,howell2011,lasky2014,rosswog2014,ravi2014,gao2016,radice2018,ai2020}, while the latter can be calculated from the total mass of the BNS-merger system extracted from the GW data. When
\begin{eqnarray}
M_{\rm rem,0} > (1+\chi_r)M_{\rm TOV},
\end{eqnarray}
a black hole (BH) or short-lived hyper-massive NS (HMNS) would be produced; When 
\begin{eqnarray}
M_{\rm rem,0} < (1+\chi_r)M_{\rm TOV},~M_{\rm b,rem} > M_{\rm b,TOV},
\end{eqnarray}
a long-lived super-massive NS (SMNS) with rigid rotation would be produced; When 
\begin{eqnarray}
M_{\rm b,rem} < M_{\rm b,TOV},
\end{eqnarray}
a stable NS (SNS) would be produced. Here $\chi_r \lesssim 0.2$ stands for the critical enhancement factor for NS mass for different types of merger remnants \citep{cook1994,lasota1996,breu2016,ai2020}. The mass with subscript ``0'' represents the value at the initial spin period for a rigidly rotating NSs, while the mass with subscript ``b'' is the baryonic mass. The remnant mass can be calculated from the total mass of the BNS system. In short, if we know both the type and mass of the merger remnant, we can obtain constraints on $M_{\rm TOV}$.

It has been long believed that BNS mergers are the source of short gamma-ray busrts (SGRBs) \citep{paczynski1986,eichler1989,paczynski1991,narayan1992}, before it was confirmed by the joint detection of GW and electromagnetic (EM) signals. Observationally, a good fraction ($\sim 24\%$, see Appendix \ref{sec:fx} for details) of sGRBs are followed by an internal X-ray plateau (a plateau followed by an extremely steep decay) on the GRB afterglow's light curve. This feature cannot be explained within the framework of external shock, but can be naturally attributed to the collapse of the post-merger SMNS to a BH. Therefore, about $24\%$ of the BNS mergers would produce a SMNS. Due to the lack of mass information on these BNS systems, no constraint on $M_{\rm TOV}$ can be made with individual sources. However, given the mass distribution of BNS merger systems, the fraction of SMNS as the merger remnants can be estimated for each $M_{\rm TOV}$. By comparing the inferred fraction from the theory with the observed one, constraints on $M_{\rm TOV}$ can be made, e.g. $M_{\rm TOV} \sim 2.3 M_{\odot}$ \citep{gao2016}.

So far, only two BNS merger events have been identified through GW observation: GW170817 \citep{abbott2017a} and GW190425 \citep{abbott2020}. GW170817 was the first event to be detected through both GW and EM observations, providing valuable insights into the physical processes involved in BNS mergers \citep{abbott2017b,abbott2017c}. However, the nature of the remnant formed after GW170817 is still a subject of debate. While the existence a short-lived hyper-massive NS after the merger can explain most of the EM signals, including the $1.7{\rm s}$ time delay for the sGRB after the merger\footnote{The $1.7s$ time delay is not necessarily due to the delayed formation of the BH. This is because: 1) Recent numerical simulations show that a jet could be launched in the BNS merger with a magnetar as the central engine \citep{kiuchi2023}; 2) The $1.7s$ delay could be decomposed in several components, with the dominant portion naturally explained by the travel time of the jet before the release of gamma-ray photons \citep{zhang2019}.} and the required ejecta mass to power the two-component kilonova, some argue that a long-lived neutron has to be produced to serve as the central engine that powers the bright kilonova emission \citep{yu2018,li2018}. Based on the scenario in which a short-lived HMNS was produced in GW170817, lasting for $\sim 1.7{\rm s}$, various studies have constrained the maximum mass of NSs \citep{margalit2017, rezzolla2018, ruiz2018}, e.g. $M_{\rm TOV} \leq 2.16M_{\odot}$ \citep{margalit2017}. To provide a comprehensive analysis, \cite{ai2020} proposed several universal relations for the masses and spin periods of rigidly rotating NSs, and summarized the constraints on $M_{\rm TOV}$ for different assumptions regarding the types of merger remnants. They claim that, if the merger remnant was a short-lived HMNS, one should have $M_{\rm TOV} < 2.09^{+0.11}_{-0.09}M_{\odot}$; if it was a long-lived SMNS, one should have $2.09^{+0.11}_{-0.09}M_{\odot}<M_{\rm TOV} < 2.43^{+0.10}_{-0.08}M_{\odot}$; if it was a SNS, one should have $M_{\rm TOV} > 2.43^{+0.10}_{-0.08}M_{\odot}$. The uncertainties are in the $2\sigma$ confidence level. The results from all of these works are under the assumption that $\chi_r \sim 0.2$. However, \cite{shibata2019} claimed that the angular momentum of the remnant might be efficiently lost during the differential rotating phase, so that the initial spin of the SMNS (if it was formed) cannot be Keplerian, and $\chi_r$ could be significantly smaller than $0.2$. Again, based on the scenario that the merger remnant of GW170817 was a short-lived HMNS, they relaxed the constraint on $M_{\rm TOV}$ to be approximately smaller than $2.3M_{\odot}$. Recently, \cite{margalit2022} also suggested that, after the merger, a rigidly rotating core inside the HMNS may form and later collapses into a BH before the differential rotation vanishes, so that $\chi_r < 0.2$.

In the case of GW190425, the total mass of the BNS system was found to be approximately $3.4M_{\odot}$ \citep{abbott2020}, which deviates significantly from the total mass of Galactic DNSs. Interestingly, a fast radio burst (FRB) FRB190425A was discovered at $\sim 2.5$ hours after the merger time of GW190425, and there is a possibility of association between the two events at a $2.8\sigma$ confidence level \citep{moroianu2023,panther2023}. If the association is indeed true, the blitzar FRB model could be applied. This model predicts that non-repeating FRBs could be emitted when a SMNS collapses into a BH \citep{zhang2014,falcke2014,most2018}. In this scenario, under the framework of NSs, roughly $M_{\rm TOV} > 2.6M_{\odot}$ is required \citep{moroianu2023}. According to the follow-up study by \cite{panther2023}, the most possible common host galaxy for GW190425 and FRB 20190425A is UGC10667, with a redshift $z = 0.03136$. Knowing the exact distance and the sky location of the source, the parameters of the BNS-merger system extracted from the GW data can be refined. \cite{bhardwaj2023} found that, if the host galaxy of GW190425 was UGC10667, our viewing angle for this BNS merger should be greater than $30$ degree, where the massive merger ejecta would cause a significant attenuation for the FRB signal at $400$MHz. Therefore, the proposed association could not be real. Note that there are still several other candidate host galaxies for FRB 20190425A that could allow the association, even though with smaller chance probabilities \citep{panther2023}. In this work, we still treat the pair as a possible association and discuss its contribution to the constraints on $M_{\rm TOV}$ and other parameters.

In this work, based on the previous constraints on NS EoS with multi-messenger method, we focus on the central remnants of BNS mergers to establish joint constraints on the maximum mass of NSs ($M_{\rm TOV}$).
Section 2 introduces the fundamental formula for the masses associated with BNS mergers. In Section 3, we gather all the observations into a Bayesian framework and employ Markov chain Monte Carlo (MCMC) simulations to infer $M_{\rm TOV}$ and other essential parameters. Our conclusions and discussions are presented in Section 5.

\section{Formula for the masses related to BNS mergers}
\subsection{remnant mass of a BNS merger}
\label{sec:Mrem}
The masses of NSs obtained from the GW data are gravitational masses, while what was conserved during the merger is the total baryonic mass of the system. In order to calculate the mass of merger remnant, universal relations that independent to NS EoSs, to convert between the gravitational mass and baryonic mass are needed. \cite{gao2020} proposed several universal relations based on a collection of common EoSs with $M_{\rm TOV}$ ranging from $2.05M_{\odot}$ to $2.78M_{\odot}$, which cover the generally believed $M_{\rm TOV}$ range. A more detailed comparison between the microscopic and macroscopic parameter spaces for the EoSs used to construct the universal relations and the parameter spaces constrained by latest multi-messenger observations is shown in Appendix \ref{sec:parameter_space}. 

Consider a BNS merger system with the masses of two pre-merger NSs as $M_1$ and $M_2$ ($M_1 \geq M_2$), respectively. 
Supposing the spins of the pre-merger NSs are relatively low, their baryonic masses can be calculated with the universal relation for non-rotating or slow-rotating NSs, which reads as \citep{gao2020}
\begin{eqnarray}
M_{b,i} = M_{i} + A_0 M_{i}^2,
\end{eqnarray}
where $i = 1,2$ and $A_0 = 0.08$. The baryonic mass of the central remnant could be calculated as
\begin{eqnarray}
M_{\rm b,rem} = M_{b,1} + M_{b,2} - M_{\rm ej},
\end{eqnarray}
where $M_{\rm ej} $ is the mass ejected during the merger 
Applying the relation of gravitational mass and baryonic mass for a post-merger rapidly rotating NS with an arbitrary rigid rotation period \citep{gao2020},
\begin{eqnarray}
M_b = M + A_s M^2
\end{eqnarray}
with $A_s = 0.073$,
the gravitational mass for the remnant $M_{\rm rem,0}$ can be estimated (assume the remnant is a rigidly rotating NS first and test the hypothesis later). Hereafter, the initial gravitational mass of the merger remnant is simply denoted as $M_{\rm rem}$ instead of $M_{\rm rem,0}$.
According to \cite{gao2020}, the conversion between the two types of masses introduce a non-negligible uncertainty, denoted as $f_{\rm err}$, which is at most $6\%$ and does not have a clear dependence on the mass of the neutron star. In this work, we add this percentage error directly to the baryonic mass of the central remnant to simplify the calculation, which reads as
\begin{eqnarray}
   M_{\rm b,rem} = M_{b,1} + M_{b,2} - M_{\rm ej} + f_{\rm err} (M_{b,1} + M_{b,2}).
\end{eqnarray}
In different converting directions, the uncertainties would partially cancel each other out, so we do not introduce another uncertainty on the gravitational mass of the merger remnant.
For GW170817, the ejecta mass inferred from the kilonova observation is $M_{\rm ej} \approx 0.06M_{\odot}$ when only radioactive heating of the ejecta is considered \citep[e.g.][]{villar2017}. In principle, the required ejecta mass could be lower if a long-lived central engine existed. The existing numerical simulations predict that the ejecta mass might be in order of $10^{-4}M_{\odot} \sim {\rm a} ~ {\rm few} \times 10^{-2}M_{\odot}$ \citep{rezzolla2011,rezzolla2013,hotokezaka2013,siegel2017,fujibayashi2018}. Generally, the gravitational mass of the central remnant can be expressed as $M_{\rm rem} = {\cal G}(M_{\rm tot}, q, f_{\rm err}, M_{\rm ej})$, where $M_{\rm tot} = M_1 + M_2$ represents the total mass and $q = M_2/M_1$ is the mass ratio, from which $M_1$ and $M_2$ can be easily calculated. For a certain total mass $M_{\rm tot}$, the influence of different mass ratios on the remnant mass is also tested. We find that, for $q = 0.5-1$, the influence is negligible. So, we always set $q = 1$ when $q$ is unknown.

\subsection{mass distribution of BNS-merger systems}
\label{sec:BNS_mass_distribution}
Currently, the measurement of the total mass of BNS merger systems can only be achieved through the detection of GWs \citep{liyinjie2021}. Due to the limited number of detections (only two events), obtaining a meaningful mass distribution for BNS merger systems is challenging. The most similar system to the BNS merger is the Galactic double neutron star (DNS) systems. Even if we exclude DNS systems that will not merge within the \emph{Hubble} timescale, the overall mass distribution remains largely unchanged \citep{farrow19}. This distribution can be described by a Gaussian distribution with $\{\mu_G, \sigma_G\} = \{2.64M_{\odot}, 0.13M_{\odot}\}$ (See Appendix \ref{sec:BNS_mass} for details). 

Before the detection of GW170817 and GW190425, it is generally believed that mass of BNS merger systems follows the mass distribution of the Galactic DNS systems. However, for GW190425, the total mass reaches $\sim 3.3M_{\odot}$ with low-spin prior assumption for the pre-merger NSs, and $\sim 3.4M_{\odot}$ with high-spin prior assumption. These values lie outside the $5\sigma$ confidence region of the distribution of the Galatic DNS systems \citep{abbott2020}. Therefore, besides the contribution of Galactic-DNS-like systems, a high-mass component must exist in the mass distribution of BNS mergers. 

\begin{figure}
	\includegraphics[width=\columnwidth]{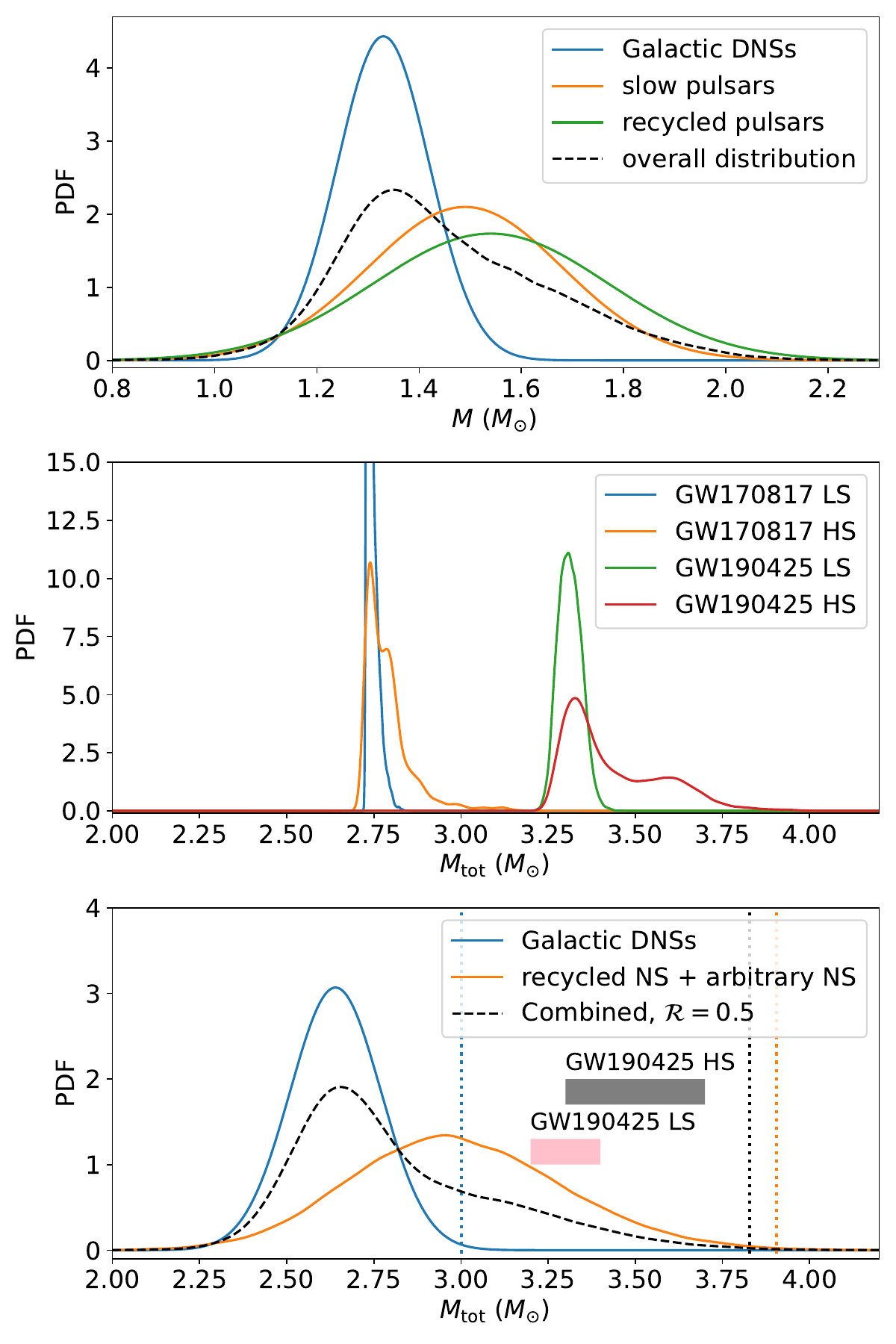}
    \caption{Upper panel: The solid lines represent the mass distributions of NSs in different populations. The black dashed line stands for the mass distribution of the secondary NSs in the high-mass BNS systems. Middle panel: The probability density distributions of the total masses ($M_{\rm tot}$) for GW170817 and GW190425, under the high-spin (HS) and low-spin priors (LS), respectively. Lower panel: The solid lines represent the mass distribution of the Galactic DNS systems and the high-mass component. The dashed black line represents the resultant mass distribution of BNS mergers with ${\cal R} = 0.5$ as an example. The dotted vertical lines represents the $3\sigma$ upper bounds for the mass distributions. The black and pink shades represent the range of the total mass for GW190425 under the low-spin and high-spin priors, respectively.}
    \label{fig:mass}
\end{figure}

Recycled pulsars, which include all millisecond pulsars and accreting NSs with low-mass companions, form a population of NSs with masses higher than their birth masses. Their mass distribution can be described by a Gaussian distribution with $\{\mu_r, \sigma_r\} = \{1.54M_{\odot}, 0.23M_{\odot}\}$, which significantly deviates from the mass distributions of slow pulsars and NSs in the Galactic DNS systems \citep{ozel2016}. The masses of slow pulsars and NSs in the Galactic DNS systems also follow Gaussian distributions, expressed as $\{\mu_s, \sigma_s\} = \{1.49M_{\odot}, 0.19M_{\odot}\}$ and $\{\mu_{\rm DNS}, \sigma_{\rm DNS}\} = \{1.33M_{\odot}, 0.09M_{\odot}\}$, respectively \citep{ozel2016}. These mass distributions are shown in the upper panel of Figure \ref{fig:mass}.  It is natural to expect that high-mass BNSs are supposed to contain at least one recycled neutron star. Given that the formation channels of merging BNS systems are still uncertain, we randomly select NSs from the three populations, each with an equal chance, following the mass distribution of each population, to form high-mass BNS systems together with the recycled NSs. The mass distribution of the secondary NSs is also shown in the upper panel of Figure \ref{fig:mass}. The overall distribution of the total masses of BNS merger systems can be written as
\begin{eqnarray}
{\cal D}_{\rm BNS}(M_{\rm tot}~|~{\cal R}) = (1 - {\cal R}){\cal D}_G(M_{\rm tot}) + {\cal R}{\cal D}_{\rm H}(M_{\rm tot}),
\end{eqnarray}
where ${\cal D}_G$ and ${\cal D}_{\rm H}$ represent the mass distributions of Galactic DNSs and the high-mass BNSs, respectively. ${\cal R}$ is a fraction parameter that represents the contribution of the high-mass component.

\begin{figure}
\resizebox{85mm}{!}{\includegraphics[]{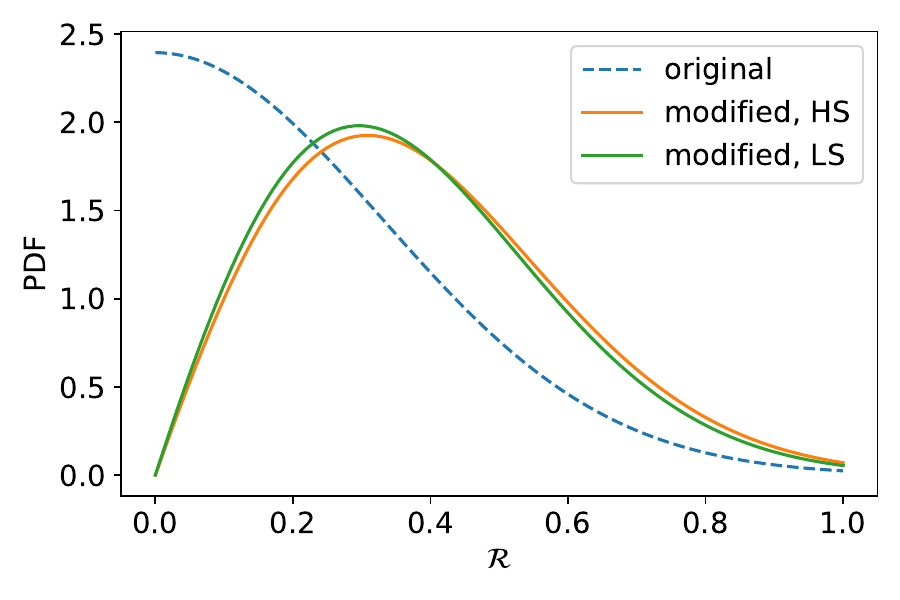}}
\caption{The probability density distributions for the fraction parameter ${\cal R}$ inferred from the total mass of GW170817 and GW190425. The measurements of $M_{\rm tot}$ for the two GW-BNS-merger events with both high-spin prior and low-spin prior assumptions are shown, respectively. We do not consider the case where one event has a high-spin prior while the other has a low-spin prior, as it could result in a curve located between the two shown modified priors.}
\label{fig:R_dis}
\end{figure}

\section{Constraints on the maximum mass of neutron stars}
Multi-messenger observations can place constraints on the maximum mass of NSs ($M_{\rm TOV}$). However, sometimes the observations are also related to other parameters, such as the maximum enhancement factor for the mass of the post-merger rigidly rotating NS ($\chi_r$), and the fraction of high-mass components for the mass distribution of BNS merger systems (${\cal R}$). Besides, we also include the ejecta mass of BNS merger ($M_{\rm ej}$) and the uncertainty of the universal relations for the $M-M_b$ conversion ($f_{\rm err}$) into the parameter set. These parameters are included in ${\bf \Theta} = \{M_{\rm TOV}, \chi_{r}, {\cal R}, f_{\rm err}, M_{\rm ej}$\}. The Bayesian inference method is applied in this section. Before constraining $M_{\rm TOV}$, we first need to determine reasonable priors for these parameters. In the absence of convincing information on $\chi_r$, it can only be assumed to be uniformly distributed in the range [$0.02$, $0.2$] \citep{shibata2019}. Similarly, the prior of $f_{\rm err}$ is set to be a uniform distribution in the range [$-6\%,6\%$] while the prior of $M_{\rm ej}$ is set to be a uniform distribution in logarithmic space in the range [$10^{-4}$, $6\times 10^{-2}$]. The probability density function of ${\cal R}$ can be inferred from the total mass of the two BNS merger systems detected via GWs (GW170817 and GW190425), independently from the other two parameters. Before the detection of BNS merger via GWs, it was assumed that the mass distribution of BNS mergers follows that for the Galatic DNSs (${\cal R} = 0.0$), while it would be highly unlikely that most of the BNS-merger systems consist of two massive millisecond pulsars (${\cal R} = 1.0$). Therefore, we set ${\cal R}$ to follow a Gaussian distribution with $\{\mu_{\cal R}, \sigma_{\cal R}\} = \{0.0,0.33\}$, ${\cal R}>0$, as the original prior ($p({\cal R})$). This distribution can be modified by the GW170817 and GW190425 systems as
\begin{eqnarray}
p({\cal R}~|~{\bf M_{\rm tot}}) \propto \left[ \prod_i \int{\cal D}_{\rm BNS}(M_{\rm tot,i}~|~{\cal R}) p(M_{\rm tot,i}) dM_{\rm tot,i}\right] p({\cal R}) 
\label{eq:R_prior}
\end{eqnarray}
where ${\bf M_{\rm tot}} = \{M_{\rm tot,i}\}$ and $M_i$ represents the total mass of each BNS-merger system. The probability density distribution ($p(M_{\rm tot,i})$) can be obtained by conducting kernel density estimation for the posteriors for the masses provided by LIGO-Virgo-Kagra science collaboration\footnote{The public data for GW170817 and GW190425 can be access through \url{https://dcc.ligo.org/LIGO-P1800370/public} and \url{https://dcc.ligo.org/LIGO-P2000026/public}, respectively.}, which are shown in the middle panel of Figure \ref{fig:mass}. The joint distributions of $M_{\rm tot}$ and mass ratio $q$ for GW170817 and GW190425 are shown in Figure \ref{fig:mass_q} in Appendix \ref{sec:Mtot_q}. The total mass distribution depends on the spin prior for the two pre-merger NSs, so that these cases will be discussed separately.
The original prior and the modified priors of ${\cal R}$ are shown in the lower panel of Figure \ref{fig:R_dis}, with the latter being used for future calculations.

We divide the observations related to $M_{\rm TOV}$ into two Categories, based on whether it can be constrained independently from the other four parameters in ${\bf \Theta}$ (Category I) or not (Category II).

\subsection{Observations: Category I}
\label{sec:category_I}

\begin{figure}
    \resizebox{80mm}{!}{\includegraphics{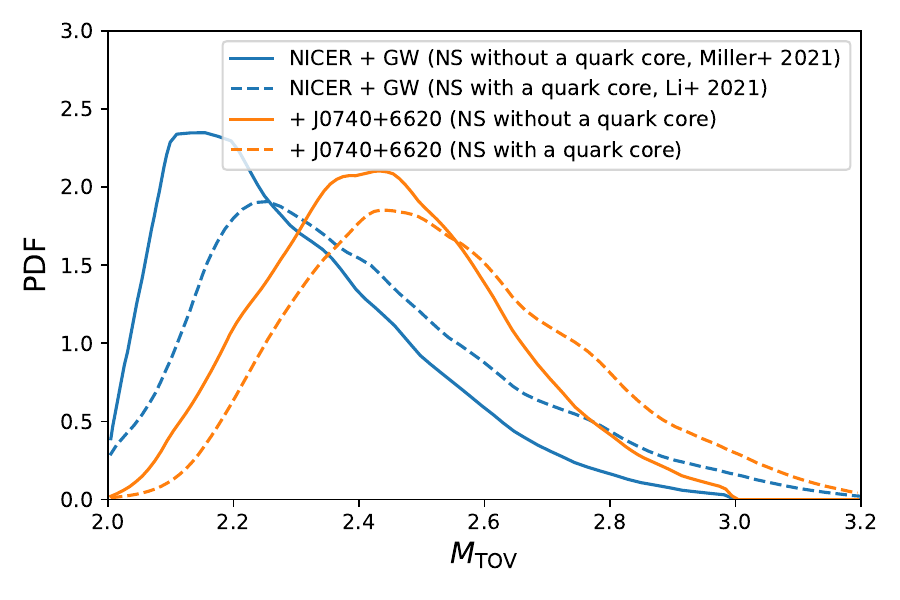}}
    \caption{Constraints on $M_{\rm TOV}$ with the facts in Catalog I. The blue and orange solid lines are extracted from \citet{miller2021} and \citet{li2021}, respectively.}
    \label{fig:Mtov_prior}
\end{figure}

The observations included in Category I are denoted as ${\bf d_{\rm I}} = \{d_{{\rm I},i}\}$, which are listed as:
\begin{itemize}
    \item $d_{{\rm I},1}$: The constraints on $M_{\rm TOV}$ presented in previous works, based on the GW data from BNS merger events GW170817 and GW190425, as well as the joint measurements for the radii and masses of some massive pulsars (i.e. PSR J0030+0451 and PSR J0740+6620), assuming the NSs are composed of predominately neutrons without a quark core \citep[i.e.][]{miller2021} or with a quark core \citep[i.e.][]{li2021}.
    
    \item $d_{{\rm I},2}$: The heaviest pulsar so far, PSR J0952-0607, with gravitational mass $M = 2.35\pm 0.17 M_{\odot}$ \citep{romani2022}. $M_{\rm TOV} > 2.35\pm 0.17 M_{\odot}$ can be obtained.
\end{itemize}
The observation $d_{{\rm I},1}$ summarizes the constraints on $M_{\rm TOV}$ from previous works and can directly give the 
probability density distribution of $M_{\rm TOV}$, which is shown in Figure \ref{fig:Mtov_prior}. When $d_{\rm I,2}$ is included, the posterior of $M_{\rm TOV}$ can be expressed as
\begin{eqnarray}
p(M_{\rm TOV}~|~{\bf d_{\rm I}}) \propto {\cal L}(d_{{\rm I},2}~|~M_{\rm TOV})p(M_{\rm TOV}~|~d_{{\rm I},1}),
\label{eq:Mtov_dI}
\end{eqnarray}
where
\begin{eqnarray}
    P(d_{{\rm I},2}~|~M_{\rm TOV}) = 
    \int_{-\infty}^{M_{\rm TOV}} \frac{1}{\sqrt{2\pi}\sigma_{\rm NS}}{\rm exp}\left[-\frac{(m-M_{\rm NS})^2}{2\sigma_{\rm NS}^2}\right] dm,
\end{eqnarray}
with $M_{\rm NS} = 2.35M_{\odot}$ and $\sigma_M = 0.17{M_\odot}$ adopted. The new PDF of $M_{\rm TOV}$ is also shown in Figure \ref{fig:Mtov_prior}. The observations in this category can be used to constrain $M_{\rm TOV}$ independently to the other two parameters. We have $M_{\rm TOV} \sim 2.43 - 2.50M_{\odot}$ with an uncertainty [$-0.19M_{\odot}$, $0.23M_{\odot}$] ([$-0.34M_{\odot}$, $0.42M_{\odot}$]) at $1\sigma$ ($2\sigma$) confidence level (see Table \ref{tab:theta_with_I&IIA}). 

\subsection{Observations: Category II}
The observations included in Category II, denoted as ${\bf d_{\rm II}} $, are related to the type of remnant of BNS mergers. They are further divided into two sub categories, IIA (${\bf d_{\rm IIA}}$) and IIB  (${\bf d_{\rm IIB}}$), based on whether they come from population properties or from one specific BNS merger. The observations in Category IIA are listed as,
\begin{itemize}
    \item $d_{{\rm IIA},1}$: According to the observation of X-ray internal plateaus in the sGRB afterglows, $\sim 24\%$ of the BNS mergers would produce a SMNS.
\end{itemize}
The observations in Category IIB are listed as,
\begin{itemize}
    \item $d_{\rm IIB,1}$: The possible association of GW190425 and FRB 20190425A indicates that the merger remnant of GW190425 might be a SMNS.
    \item $d_{\rm IIB,2}$: The type of merger remnant of GW170817 is under debate. In different cases, $d_{\rm IIB,2}$ can be written as $d_{\rm IIB,2,BH}$, $d_{\rm IIB,2,SMNS}$ and $d_{\rm IIB,2,SNS}$, respectively.
    
\end{itemize}
Generally, the posterior of  ${\bf \Theta}$ can be expressed as 
\begin{eqnarray}
p({\bf \Theta} ~|~ {\bf d_{\rm II}}) &\propto&
 \left[\prod_i {\cal L}(d_{\rm IIA,i}~|~{\bf \Theta})\right] \left[\prod_i {\cal L}(d_{\rm IIB,i}~|~{\bf \Theta})\right] \nonumber \\
 &\times& p(M_{\rm TOV}~|~{\bf d_{\rm I}})p(\chi_r)p({\cal R}~|~{\bf M}).
\end{eqnarray}
where the priors of $M_{\rm TOV}$ and ${\cal R}$ in this step come from Equations \ref{eq:R_prior} and \ref{eq:Mtov_dI}, respectively.

In details, for the observation $d_{\rm IIA,1}$, the likelihood can be expressed as
\begin{eqnarray}
{\cal L}(d_{\rm IIA,1}~|~{\bf \Theta}) \propto p(f_{\rm SMNS} ~|~ M_{\rm TOV}, \chi_r, {\cal R}, f_{\rm err}, M_{\rm ej}, f_X),
\end{eqnarray}
where $f_X = 0.243$ (see Appendix \ref{sec:fx}) represents the observational fraction of X-ray internal plateau in sGRBs. According to the De Moivre-Laplace Central Limit Theorem, we know that the probability of the observed fraction deviates from the real fraction within a small positive value $\epsilon$ is
\begin{eqnarray}
    P(|f_X - f_{\rm SMNS}| < \epsilon) \approx 2\Phi \left(\epsilon \sqrt{\frac{n}{f_{\rm SMNS}(1-f_{\rm SMNS})}}\right) - 1,
\label{eq:central_value_theorem}
\end{eqnarray}
where $\Phi (x)$ is the cumulative distribution function for the standard normal distribution and $n = 144$ is the total number of sGRBs in our sample (see Appendix \ref{sec:fx}). The approximation in Equation \ref{eq:central_value_theorem} is valid when $nf_{\rm SMNS} > 5$. Therefore, the likelihood can be further written as
\begin{eqnarray}
    {\cal L}(d_{\rm IIA,1}~|~{\bf \Theta}) \propto \frac{1}{\sqrt{2\pi}} {\rm exp}\left\{-\frac{\left[\epsilon^{\prime}\sqrt{\frac{n}{f_{\rm SMNS}(1 - f_{\rm SMNS})}}\right]^2}{2}\right\}
\label{eq:likelihood_dIIA_1}
\end{eqnarray}
where
\begin{eqnarray}
    \epsilon^{\prime} = |f_X - f_{\rm SMNS}({\bf \Theta})|.
\label{eq:likelihood_dIIA_2}
\end{eqnarray}
For a certain parameter set ${\bf \Theta}$, the predicted fraction of SMNS as the merger remnant can be written as
\begin{eqnarray}
    f_{\rm SMNS} = \frac{\int_{M_{\rm th,1}}^{M_{\rm th,2}}p(M_{\rm rem}~|~{\cal R},q,f_{\rm err}, M_{\rm ej})dM_{\rm rem}}{\int_{0}^{\infty}p(M_{\rm rem}~|~{\cal R},q,f_{\rm err}, M_{\rm ej})dM_{\rm rem}},
\end{eqnarray}
where 
\begin{eqnarray}
    M_{\rm th,1} = (1 + \chi_r) M_{\rm TOV}
\end{eqnarray}
and
\begin{eqnarray}
    M_{\rm th,2} = \frac{-1 + \sqrt{1+4A_s M_{\rm b,TOV}}}{2A_s}.
\end{eqnarray}
The probability density distribution of $M_{\rm rem}$ under a certain parameter set can be expressed as
\begin{eqnarray}
    &p(M_{\rm rem}(M_{\rm tot},q, f_{\rm err}, M_{\rm ej})~|~ {\cal R}, q, f_{\rm err}, M_{\rm ej}) dM_{\rm rem}& \\ \nonumber 
    &\propto {\cal D}(M_{\rm tot}~|~{\cal R})dM_{\rm tot},&
\end{eqnarray}
where
\begin{eqnarray}
    dM_{\rm rem} = \frac{\partial {\cal G}(M_{\rm tot},q,f_{\rm err}, M_{\rm ej})}{\partial M_{\rm tot}} dM_{\rm tot}.
\end{eqnarray}


For the observations ${\bf d_{\rm IIB,1}}$, the likelihood function can be expressed as
\begin{eqnarray}
{\cal L}(d_{\rm IIB,1}~|~{\bf \Theta}) = \int dM_{\rm tot} \int dq {\cal L}(d_{\rm IIB,1}~|~{\bf \Theta}, M_{\rm tot}, q)p(M_{\rm tot},q),
\end{eqnarray}
where
\begin{eqnarray}
  {\cal L}(d_{\rm IIB,1}~|~{\bf \Theta}, M_{\rm tot}, q) = \left\{ \begin{array}{cc}
     1  & {\rm if}~{\rm GW190425-SMNS}, \\
     0  & {\rm ELSE},
  \end{array} \right.
\end{eqnarray}
and $p(M_{\rm tot},q)$ can be numerically obtained by conducting 2D kernel density estimation from the posteriors of $M_{\rm tot}$ and $q$ from GW public data (See detailed in Appendix \ref{sec:Mtot_q}). ${\cal L}(d_{\rm IIB,2}~|~{\bf \Theta})$ for GW170817 can be expressed in a similar way but different types of merger remnant are considered. For each parameter set, the merger remnant can be determined following the criteria listed in the introduction.

\subsection{Results}
\begin{figure*}
    \centering
    \resizebox{160mm}{!}{\includegraphics{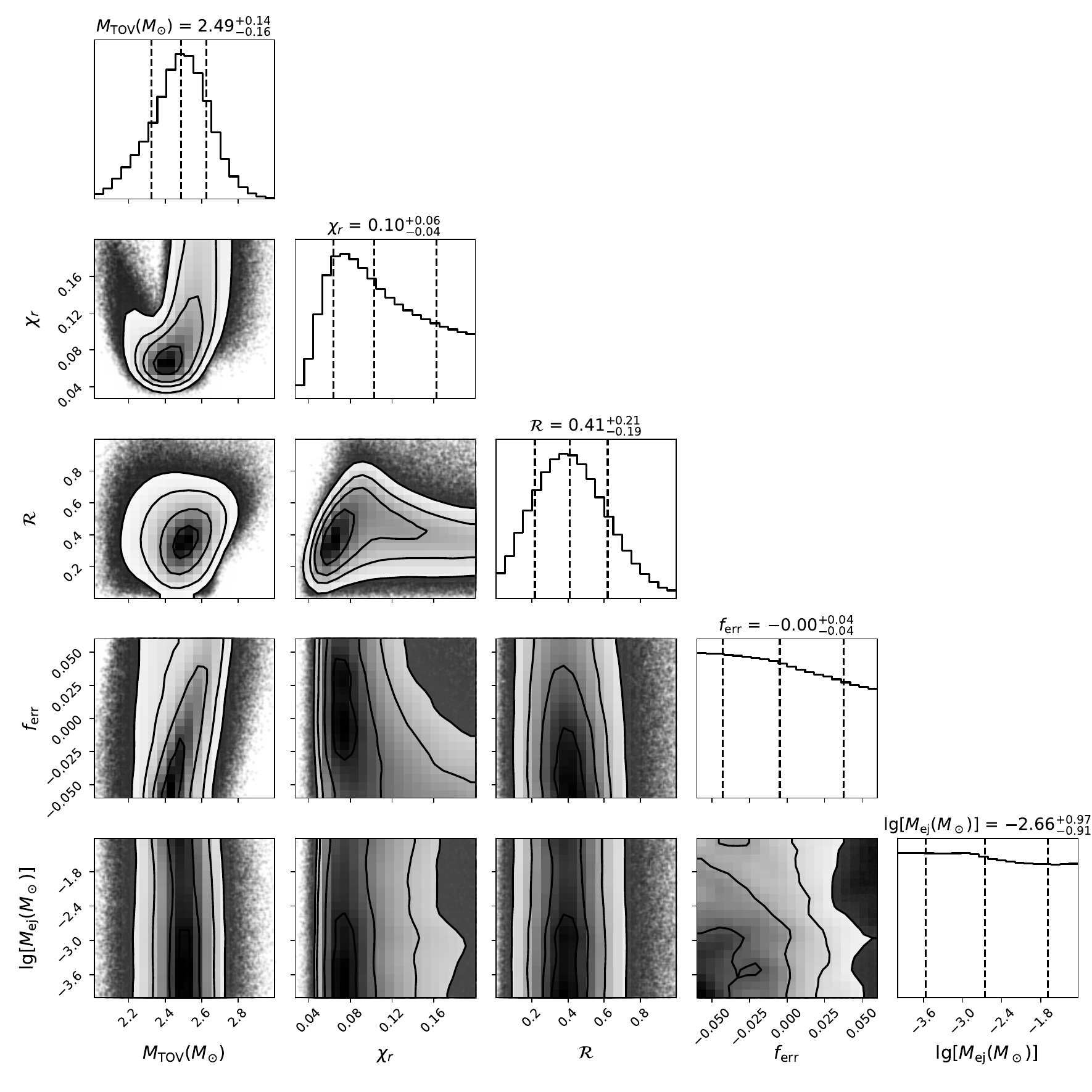}}
    \caption{Constraints on ${\bf \Theta}$ with the facts in Category I and IIA. Low-spin priors were adopted when using the masses of the BNSs GW170817 and GW190425, to estimate the prior of ${\cal R}$. The neutron stars are assumed to be composed of predominately neutrons without a quark core. The titles show the central value and uncertainties at $1\sigma$ confidence level.}
    \label{fig:theta_posterior}
\end{figure*}

We employ the python module \emph{emcee} \citep{Foreman-Mackey2013} to conduct the Bayesian MCMC parameter estimation. The constraints on $M_{\rm TOV}$ solely with ${\bf d_{\rm I}}$ has been presented in Section \ref{sec:category_I}. Here, we first combine the observations in Category I and IIA to constrain ${\bf \Theta}$. Figure \ref{fig:theta_posterior} shows an example with $M_{\rm tot}$ inferred using low-spin prior for both GW170817 and GW190425, assuming the NSs are made of pure neutrons. It is worth noting that, instead of close to $0.2$, the probability density distribution of $\chi_r$ is supposed to peak at about $0.1$, although with a relatively large uncertainty. The detailed results with different assumptions are listed in Table \ref{tab:theta_with_I&IIA}. In general, when considering both ${\bf d_{\rm I}}$ and ${\bf d_{\rm IIA}}$, we find $M_{\rm TOV} \approx 2.49M_{\odot} - 2.52M_{\odot}$ with an uncertainty range of [$-0.16M_{\odot}$, $0.15 M_{\odot}$] ([$-0.28M_{\odot}$, $0.26 M_{\odot}$]); $\chi_r \approx 0.10$ with an uncertainty range of [$-0.04$, $0.06$] ([$-0.06$, $0.08$]); and ${\cal R} \approx 0.41 - 0.44$ with an uncertainty range of [$-0.20$, $0.22$] ([$-0.31$, $0.36$]). All the uncertainties presented here are at $1\sigma$ ($2\sigma$) confidence level.

\begin{table*}
\caption{Constraints on ${\bf \Theta}$ with observations in Category I (${\bf d_{\rm I}}$, including the measurements for compactness, tidal deformabilities and the mass of the most massive pulsar) and IIA (${\bf d_{\rm IIA}}$, the fraction of X-ray internal plateaus in sGRBs). ``HS" and ``LS" represent the high-spin prior and low-spin prior for the pre-merger NSs in GW170817 and GW190425, which influence the modified prior of ${\cal R}$.}
\centering
\begin{tabular}{|c|c|c|c|}
  \hline
  &$\makecell[c]{~\\{\bf d_{\rm I}}\\~} $ & ${\bf d_{\rm I}} + {\bf d_{\rm IIA}}$  \\\hline
\makecell[c]{NS without \\ a quark core\\ (HS)} & \makecell[c]{$M_{\rm TOV} = 2.43^{+0.20(+0.41)}_{-0.18(-0.32)}M_{\odot}$} & \makecell[c]{~\\$M_{\rm TOV} = 2.49^{+0.14(+0.23)}_{-0.16(-0.28)}M_{\odot}$\\~\\$\chi_r = 0.10^{+0.06(+0.08)}_{-0.05(-0.05)}$\\~\\${\cal R} = 0.42^{+0.22(+0.36)}_{-0.20(-0.30)}$\\~}  \\ \hline
\makecell[c]{NS without \\ a quark core\\ (LS)} & ---- & \makecell[c]{~\\$M_{\rm TOV} = 2.49^{+0.14(+0.23)}_{-0.16(-0.28)}M_{\odot}$\\~\\$\chi_r = 0.10^{+0.06(+0.09)}_{-0.04(-0.05)}$\\~\\${\cal R} = 0.41^{+0.21(+0.36)}_{-0.19(-0.29)}$\\~}  \\ \hline
\makecell[c]{NS with \\ a quark core\\ (HS)} & \makecell[c]{$M_{\rm TOV} = 2.50^{+0.23(+0.42)}_{-0.19(-0.34)}M_{\odot}$} & \makecell[c]{~\\$M_{\rm TOV} = 2.52^{+0.15(+0.26)}_{-0.16(-0.28)}M_{\odot}$\\~\\$\chi_r = 0.11^{+0.06(+0.08)}_{-0.04(-0.06)}$\\~\\${\cal R} = 0.44^{+0.21(+0.36)}_{-0.20(-0.31)}$\\~} \\ \hline
\makecell[c]{NS with \\ a quark core\\ (LS)} & ---- & \makecell[c]{~\\$M_{\rm TOV} = 2.52^{+0.15(+0.25)}_{-0.16(-0.28)}M_{\odot}$\\~\\$\chi_r = 0.11^{+0.06(+0.08)}_{-0.04(-0.06)}$\\~\\${\cal R} = 0.42^{+0.21(+0.36)}_{-0.20(-0.30)}$\\~} \\ \hline
\end{tabular}
\label{tab:theta_with_I&IIA}
\end{table*}

\begin{figure*}
    \centering
    \begin{tabular}{ll}
    \resizebox{80mm}{!}{\includegraphics{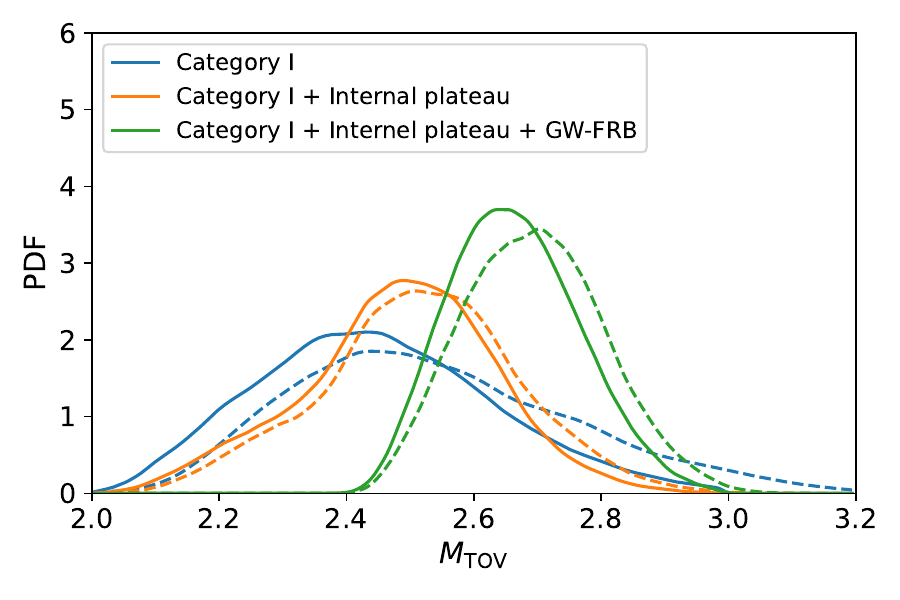}} &
    \resizebox{80mm}{!}{\includegraphics{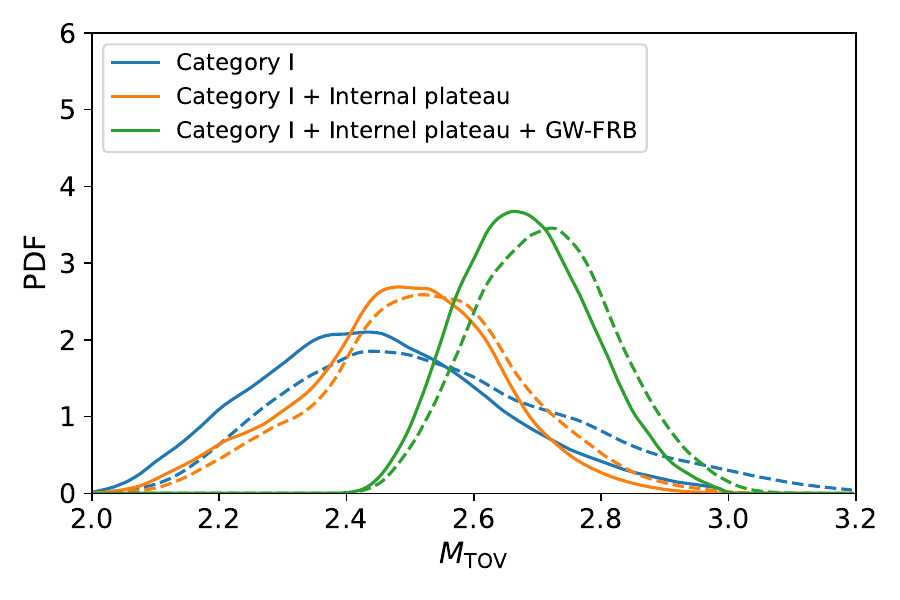}}
    \\
    \resizebox{80mm}{!}{\includegraphics{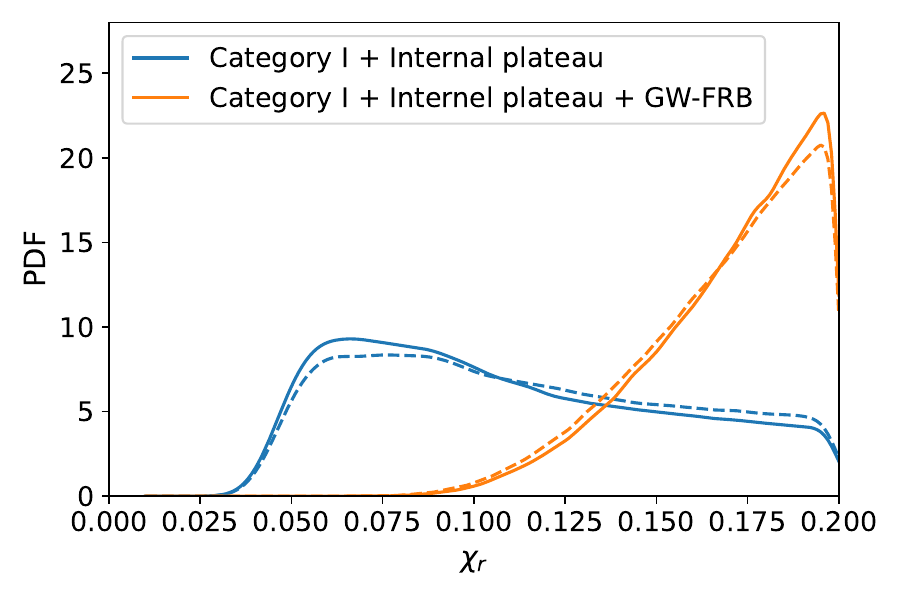}} &
    \resizebox{80mm}{!}{\includegraphics{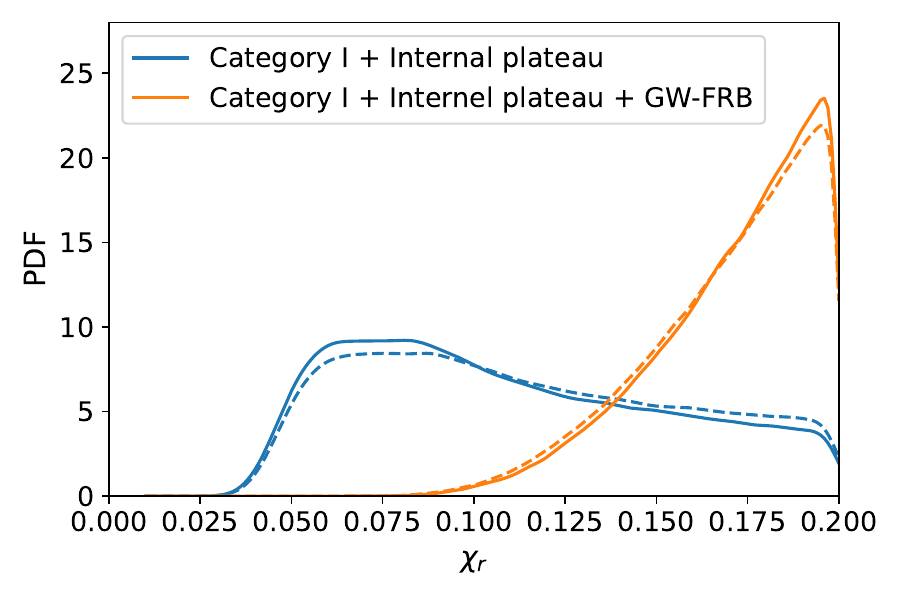}}
    \\
    \resizebox{80mm}{!}{\includegraphics{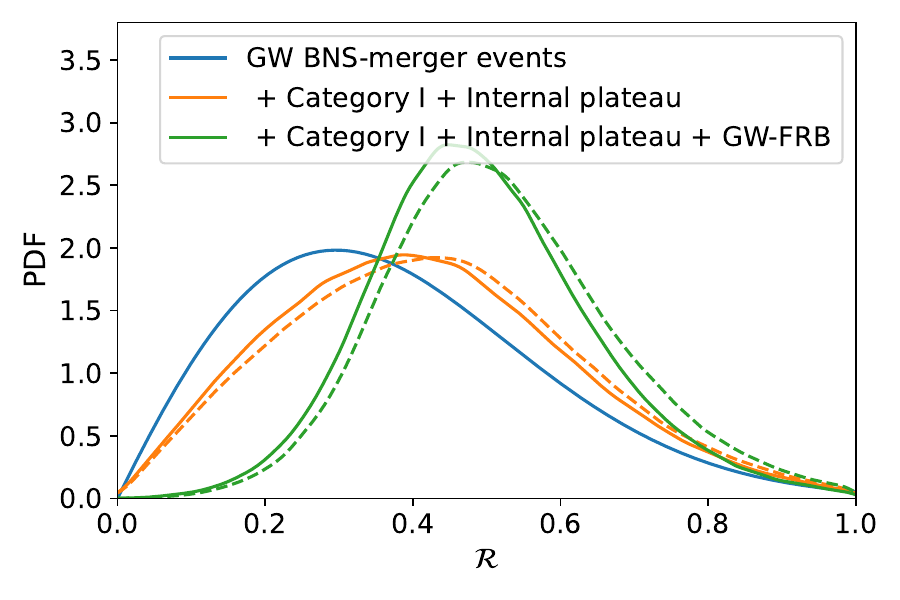}} &
    \resizebox{80mm}{!}{\includegraphics{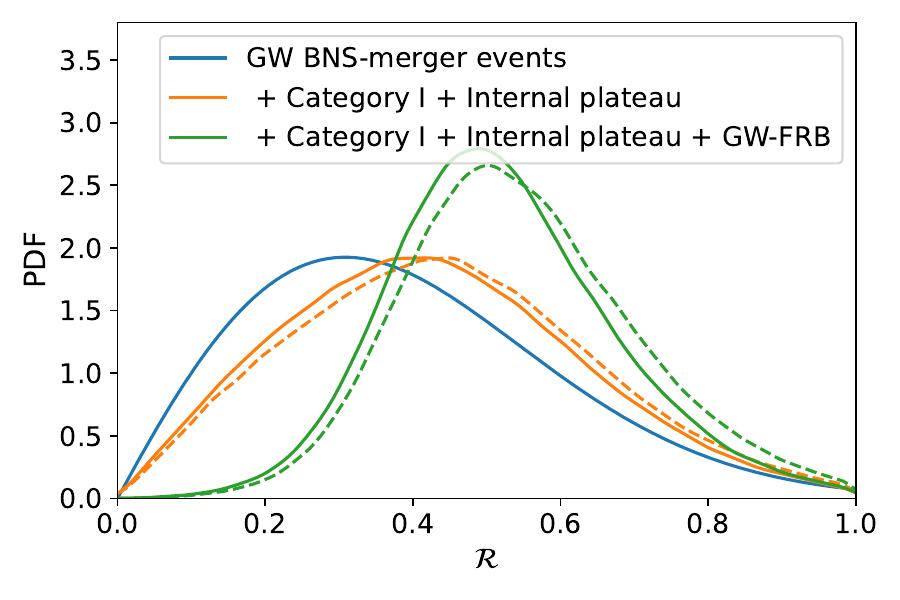}}
    \end{tabular}
    \caption{Posteriors of ${\bf \Theta} = \{M_{\rm TOV}, \chi_r, {\cal R}\}$ with observations ${\bf d_{\rm I}}$, ${\bf d_{\rm IIA}}$ and $d_{\rm IIB,1}$. In each panel, solid lines and dashed line represent the scenarios that NSs without and with a quark core, respectively. The left panel shows the case that low-spin priors are adopted for the pre-merger NSs in both GW170817 and GW190425, while the right panel shows the high-spin-prior case.}
    \label{fig:constraints_IIB_190425}
\end{figure*}

\begin{figure*}
    \centering
    \begin{tabular}{ll}
    \resizebox{80mm}{!}{\includegraphics{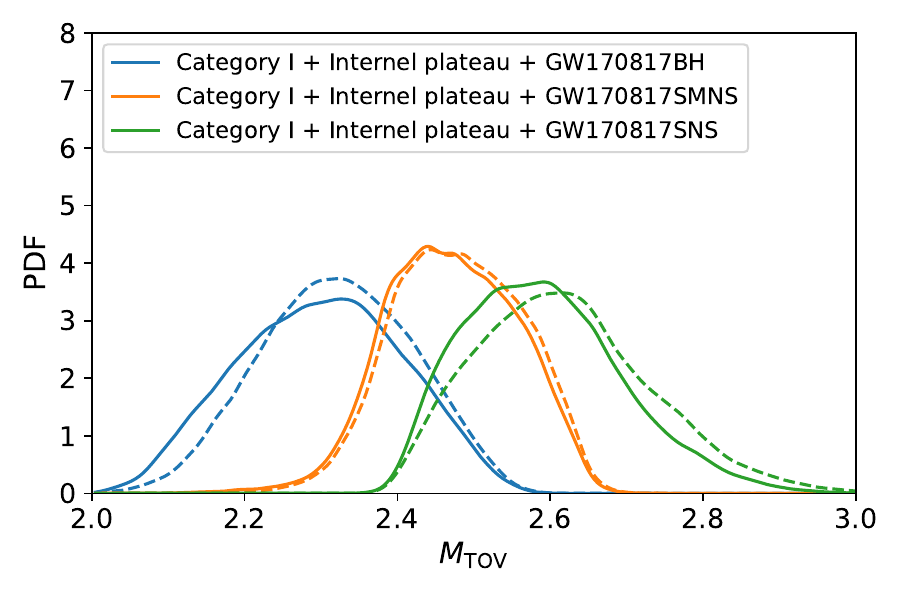}} &
    \resizebox{80mm}{!}{\includegraphics{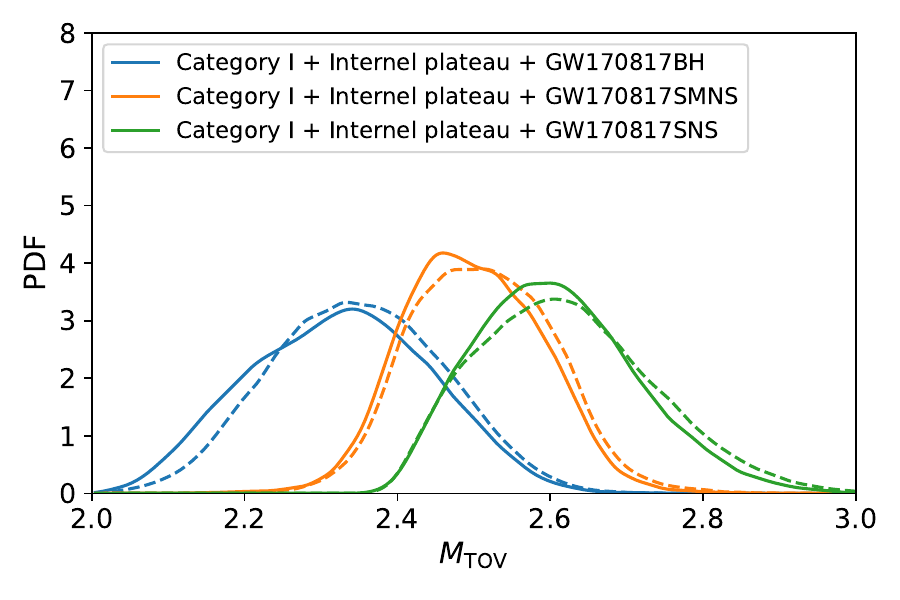}}
    \\
    \resizebox{80mm}{!}{\includegraphics{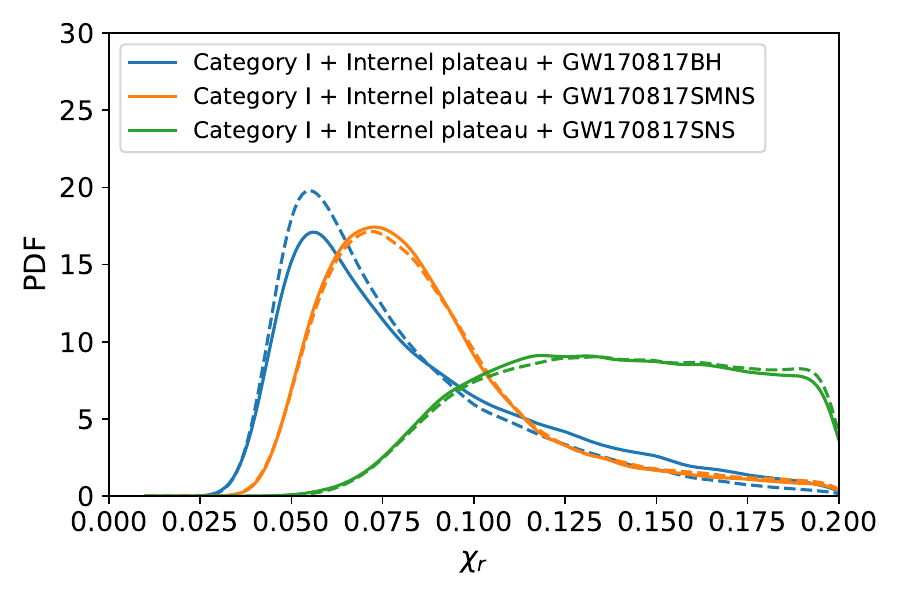}} &
    \resizebox{80mm}{!}{\includegraphics{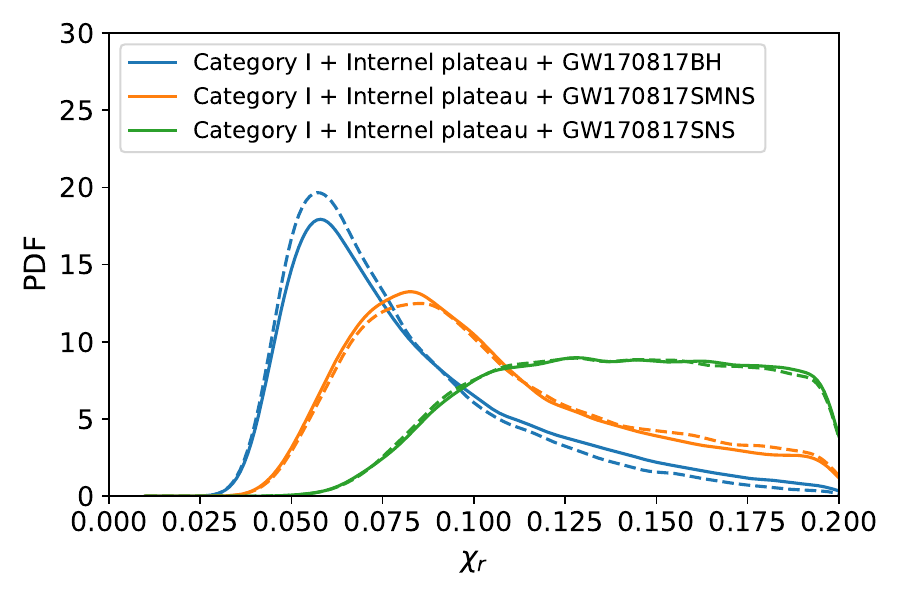}}
    \\
    \resizebox{80mm}{!}{\includegraphics{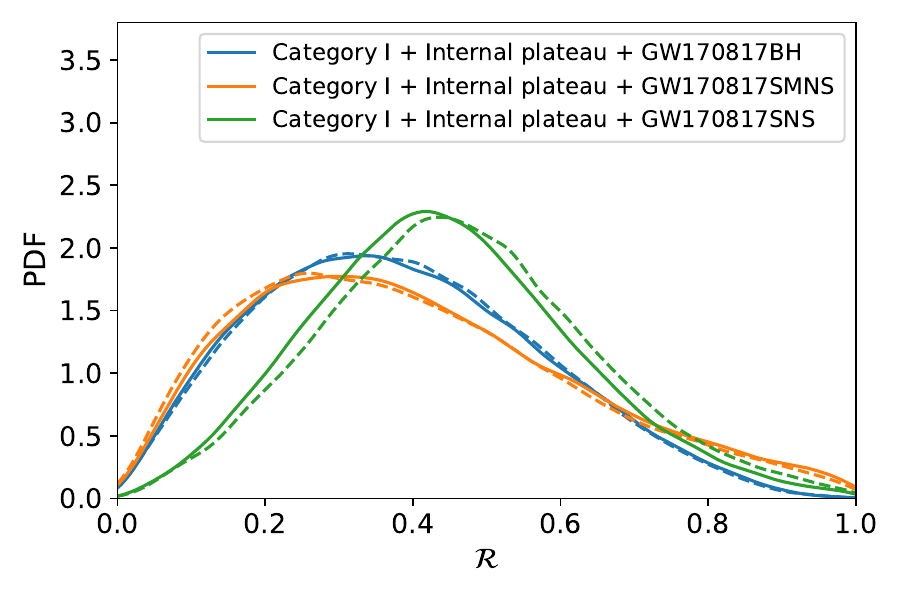}} &
    \resizebox{80mm}{!}{\includegraphics{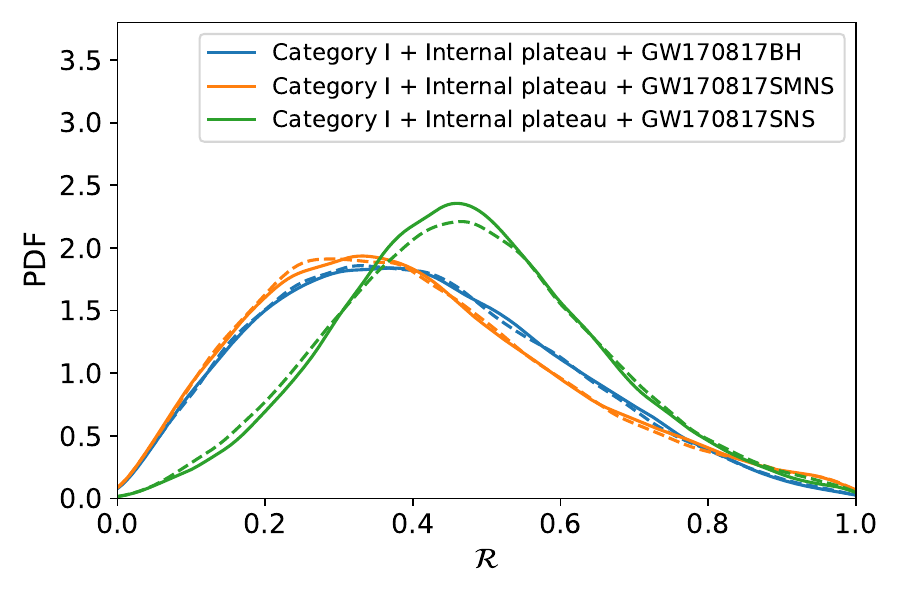}}
    \end{tabular}
    \caption{Similar as Figure \ref{fig:constraints_IIB_190425}, but with observations ${\bf d_{\rm I}}$, ${\bf d_{\rm IIA}}$ and $d_{\rm IIB,2}$.}
    \label{fig:constraints_IIB_170817}
\end{figure*}

The observations in Category IIB are not confirmed, so that we have to discuss them case by case. 
Generally, the posterior of  ${\bf \Theta}$ is expressed as 
\begin{eqnarray}
p({\bf \Theta}~|~d_{\rm IIB,i},{\bf d_{\rm IIA}}, {\bf d_{\rm I}}) &\propto& {\cal L}(d_{\rm IIB,i}~|~{\bf \Theta})p({\bf \Theta}~|~{\bf d_{\rm IIA}}, {\bf d_{\rm I}}) \nonumber \\
&\propto& {\cal L}(d_{\rm IIB,i}~|~{\bf \Theta}){\cal L}({\bf d_{\rm IIA}~|~{\bf \Theta}})p({\bf \Theta}~|~{\bf d_{\rm I}}).
\end{eqnarray}
The posteriors of ${\bf \Theta}$ with observations in Category I and IIA, along with one certain case in Category IIB are shown in Figure \ref{fig:constraints_IIB_190425} and \ref{fig:constraints_IIB_170817}. These figures display the inferences made using the remnant type of GW190425 and GW170817, respectively. The central values and $1\sigma$ ($2\sigma$) uncertainties with different assumptions are listed in Table \ref{tab:theta_with_IIB}. The results can be summarized as follows:
\begin{itemize}
    \item If one assumes that the remnant of GW170817 collapses directly into a BH after the differential rotating phase, a preferred value of $M_{\rm TOV}$ around $2.30M_{\odot} - 2.35M_{\odot}$ is obtained, with an uncertainty range of [$-0.13M_{\odot}$, $0.12M_{\odot}$] ([$-0.20M_{\odot}$,$0.19M_{\odot}$]) at the $1\sigma$ ($2\sigma$) confidence level. The maximum enhancement factor for the mass of a rigidly rotating NS produced after a BNS merger would be more likely to be $\chi_r \sim 0.07$ with an uncertainty range of [$-0.03$,$0.05$] ([$-0.03$, $0.08$]) at the $1\sigma$ ($2\sigma$) confidence level. 

    \item If one assumes that the remnant of GW170817 is a SMNS, a preferred value of $M_{\rm TOV}$ around $2.47 M_{\odot} - 2.51M_{\odot}$ is obtained, with an uncertainty range of [$-0.09M_{\odot}$, $0.10M_{\odot}$] ([$-0.14M_{\odot}$, $0.15M_{\odot}$]) at the $1\sigma$ ($2\sigma$) confidence level. The maximum enhancement factor for the mass of a rigidly rotating NS produced after a BNS merger would be more likely to be $\chi_r \sim 0.08 - 0.10$ with an uncertainty range of [$-0.03$, $0.06$] ([$-0.04$, $0.08$]) at the $1\sigma$ ($2\sigma$) confidence level. 

    \item If one assumes that the remnant of GW170817 is a SNS, a preferred value of $M_{\rm TOV}$ around $2.58 M_{\odot} - 2.63M_{\odot}$ is obtained, with an uncertainty range of [$-0.10M_{\odot}$, $0.12M_{\odot}$] ([$-0.17M_{\odot}$, $0.21M_{\odot}$]) at the $1\sigma$ ($2\sigma$) confidence level. The maximum enhancement factor for the mass of a rigidly rotating NS produced after a BNS merger would be more likely to be $\chi_r \sim 0.14$ with an uncertainty range of [$-0.04$, $0.04$] ([$-0.06$, $0.05$]) at the $1\sigma$ ($2\sigma$) confidence level. 

    \item If the association of GW190425 and FRB 20190425A is real, a preferred value of $M_{\rm TOV}$ around $2.66M_{\odot} - 2.69M_{\odot}$ is obtained, with an uncertainty range of [$-0.11M_{\odot}$, $0.11M_{\odot}$]([$-0.17M_{\odot}$, $0.19M_{\odot}$]) at the $1\sigma$ ($2\sigma$) confidence level. In this case, it is highly likely that the merger remnant of GW170817 is a SNS. The value of $\chi_r$ should be very close to $0.2$.
\end{itemize}
In all cases, the fraction of high-mass component in the BNS mass distribution, denoted as ${\cal R}$, cannot be well constrained, while the central value is around ${\cal R} \sim 0.37 - 0.51$.

\begin{table*}
\caption{Constraints on ${\bf \Theta}$ with observations in Category I, IIA, and IIB (${\bf d_{\rm IIB}}$, remnant type in GW170817 or G190425) with uncertainty at $1\sigma$ ($2\sigma$) confidence level. ``HS" and ``LS" have the same definitions as in Table \ref{tab:theta_with_IIB}}
\centering
\begin{tabular}{|c|c|c|c|c|c|}
  \hline
 \backslashbox[30mm]{}{\makecell[c]{~\\$d_{\rm IIB,i}$\\~}} & \makecell[c]{~\\GW170817-BH\\~}  & GW170817-SMNS  & GW170817-SNS & GW190425-FRB 20190425A  \\\hline
\makecell[c]{NS without \\ a quark core\\ (HS)} & \makecell[c]{~\\$M_{\rm TOV} = 2.33^{+0.12(+0.19)}_{-0.13(-0.20)}M_{\odot}$\\~\\$\chi_r = 0.07^{+0.05(+0.08)}_{-0.02(-0.03)}$\\~\\${\cal R} = 0.39^{+0.23(+0.37)}_{-0.19(-0.28)}$ \\~} & \makecell[c]{~\\$M_{\rm TOV}=2.49^{+0.10(+0.15)}_{-0.09(-0.13)}M_{\odot}$\\~\\$\chi_r = 0.10^{+0.06(+0.08)}_{-0.03(-0.04)}$\\~\\${\cal R}=0.37^{+0.24(+0.42)}_{-0.18(-0.27)}$\\~} & \makecell[c]{$M_{\rm TOV}=2.60^{+0.11(+0.19)}_{-0.10(-0.15)}M_{\odot}$\\~\\$\chi_r = 0.14^{+0.04(+0.05)}_{-0.04(-0.06)}$\\~\\${\cal R}= 0.47^{+0.18(+0.32)}_{-0.17(-0.27)}$} & \makecell[c]{~\\$M_{\rm TOV} = 2.68^{+0.11(+0.18)}_{-0.10(-0.16)}M_{\odot}$\\~\\$\chi_r = 0.18^{+0.02(+0.02)}_{-0.03(-0.05)}$\\~\\${\cal R} = 0.51^{+0.16(+0.29)}_{-0.13(-0.22)}$ \\~}\\ \hline
\makecell[c]{NS without \\ a quark core\\ (LS)} & \makecell[c]{~\\$M_{\rm TOV} = 2.30^{+0.11(+0.17)}_{-0.12(-0.18)}M_{\odot}$\\~\\$\chi_r = 0.07^{+0.05(+0.09)}_{-0.03(-0.03)}$\\~\\${\cal R} = 0.37^{+0.22(+0.35)}_{-0.18(-0.27)}$\\~} & \makecell[c]{~\\$M_{\rm TOV} = 2.47^{+0.09(+0.14)}_{-0.08(-0.13)}M_{\odot}$\\~\\$\chi_r = 0.08^{+0.03(+0.07)}_{-0.02(-0.03)}$\\~\\${\cal R} = 0.37^{+0.26(+0.44)}_{-0.20(-0.28)}$\\~} & \makecell[c]{~\\$M_{\rm TOV} = 2.58^{+0.11(+0.20)}_{-0.10(-0.14)}M_{\odot}$\\~\\$\chi_r = 0.14^{+0.04(+0.05)}_{-0.04(-0.06)}$\\~\\${\cal R} = 0.43^{+0.18(+0.33)}_{-0.17(-0.27)}$\\~} & \makecell[c]{~\\$M_{\rm TOV} = 2.66^{+0.11(+0.19)}_{-0.10(-0.17)}M_{\odot}$\\~\\$\chi_r=0.18^{+0.02(+0.02)}_{-0.03(-0.05)}$\\~\\${\cal R} = 0.48^{+0.16(+0.28)}_{-0.13(-0.22)}$\\~} \\ \hline
\makecell[c]{NS with \\ a quark core\\ (HS)} & \makecell[c]{~\\$M_{\rm TOV} = 2.35^{+0.12(+0.19)}_{-0.11(-0.18)}M_{\odot}$\\~\\$\chi_r = 0.07^{+0.04(+0.07)}_{-0.02(-0.03)}$\\~\\${\cal R} = 0.39^{+0.23(+0.38)}_{-0.19(-0.28)}$\\~} & \makecell[c]{~\\$M_{\rm TOV}=2.51^{+0.10(+0.15)}_{-0.09(-0.14)}M_{\odot}$\\~\\$\chi_r = 0.10^{+0.05(+0.08)}_{-0.03(-0.04)}$\\~\\${\cal R} = 0.37^{+0.24(+0.42)}_{-0.18(-0.27)}$\\~} & \makecell[c]{$M_{\rm TOV}=2.63^{+0.12(+0.21)}_{-0.11(-0.17)}M_{\odot}$\\~\\$\chi_r = 0.14^{+0.04(+0.05)}_{-0.04(-0.06)}$\\~\\${\cal R} = 0.49^{+0.18(+0.32)}_{-0.17(-0.28)}$} & \makecell[c]{~\\$M_{\rm TOV} = 2.71^{+0.11(+0.19)}_{-0.11(-0.17)}M_{\odot}$\\~\\$\chi_r=0.17^{+0.02(+0.02)}_{-0.03(-0.05)}$\\~\\${\cal R} = 0.53^{+0.17(+0.29)}_{-0.14(-0.23)}$\\~} \\ \hline
\makecell[c]{NS with \\ a quark core\\ (LS)} & \makecell[c]{~\\$M_{\rm TOV} = 2.32^{+0.10(+0.16)}_{-0.10(-0.16)}M_{\odot}$\\~\\$\chi_r = 0.07^{+0.04(+0.08)}_{-0.02(-0.03)}$\\~\\${\cal R} = 0.37^{+0.21(+0.44)}_{-0.18(-0.28)}$\\~} & \makecell[c]{~\\$M_{\rm TOV} = 2.48^{+0.09(+0.13)}_{-0.08(-0.13)}M_{\odot}$\\~\\$\chi_r = 0.08^{+0.03(+0.08)}_{-0.02(-0.03)}$\\~\\${\cal R} = 0.37^{+0.26(+0.44)}_{-0.20(-0.28)}$\\~} & \makecell[c]{~\\$M_{\rm TOV} = 2.61^{+0.12(+0.21)}_{-0.11(-0.16)}M_{\odot}$\\~\\$\chi_r = 0.14^{+0.04(+0.05)}_{-0.04(-0.06)}$\\~\\${\cal R} = 0.45^{+0.19(+0.33)}_{-0.17(-0.28)}$\\~} & \makecell[c]{~\\$M_{\rm TOV} = 2.69^{+0.11(+0.19)}_{-0.11(-0.17)}M_{\odot}$\\~\\$\chi_r=0.17^{+0.02(+0.02)}_{-0.03(-0.05)}$\\~\\ ${\cal R}=0.51^{+0.17(+0.29)}_{-0.14(-0.22)}$\\~} \\ \hline
\end{tabular}
\label{tab:theta_with_IIB}
\end{table*}



\section{Conclusion and Discussion}
We attempted to constrain the maximum mass of NSs ($M_{\rm TOV}$), the maximum enhancement factor for the mass of rigidly rotating NSs produced after BNS mergers ($\chi_r$), and the fraction of high-mass components in the BNS merger mass distribution (${\cal R}$) by combining as many astronomical observations as possible.

Starting from the previous constraints on $M_{\rm TOV}$ in the literature based on the compactness (tidal deformability) measurements for NSs, we added constraints on $M_{\rm TOV}$ from the most massive pulsars observed so far (PSR J0952-0607), and the facts that approximately $24\%$ sGRBs are followed by an X-ray internal plateau, which indicates that about $24\%$ of BNS mergers would produce a SMNS. This led to a preferred value of $M_{\rm TOV}$ to be approximately $2.49M_{\odot} - 2.52M_{\odot}$, with an uncertainty of [$-0.16M_{\odot}$, $0.15M_{\odot}$]([$-0.28M_{\odot}$, $0.26M_{\odot}$]) at the 1$\sigma$ (2$\sigma$) confidence level.

Furthermore, we can impose additional constraints based on the type of merger remnants in GW170817 and GW190425. The remnant type could be inferred from the electromagnetic signals during and after the merger, but is still under debate. If the remnant collapsed into a BH immediately after the differential rotating phase, a value of $M_{\rm TOV}$ around $2.3M_{\odot} - 2.4M_{\odot}$ would be preferred. In this case, the maximum mass of rigidly rotating NSs produced by BNS mergers can only be $\chi_r \sim 0.07$ greater than $M_{\rm TOV}$; If the remnant was a SMNS, a higher value of $M_{\rm TOV}$ around $2.5M_{\odot}$ would be preferred. The central value of $\chi_r$ would be around $0.08$ - $0.10$; If the merger remnant was a SNS, a larger value of $M_{\rm TOV}$ around $2.6M_{\odot}$ would be preferred. The central value of $\chi_r$ would be approximately $0.14$. The constraints on ${\cal R}$, the fraction of high-mass components in the BNS merger mass distribution, are relatively loose. However, in any case, a high mass component for the mass distribution of BNS merger system is expected to exist.

The possible association of GW190425 and FRB 20190425A indicates that the merger remnant of GW190425 might be a SMNS. If that was true, $M_{\rm TOV} \sim 2.7M_{\odot}$ would be preferred while $\chi_r$ should be very close to $0.2$. Again, the constraints on ${\cal R}$ is loose.


If a SNS was produced in GW170817, constraints from the observations of its electromagnetic counterparts suggest that the surface magnetic field of the remnant should be relatively low, with a poloidal magnetic field $B_p$ less than approximately $10^{11}$ G \citep{ai2018}. However, numerical simulations indicate that the magnetic field on the surface of neutron stars can be highly amplified during the merger \citep[e.g.][]{kiuchi2023}. Although most simulations only cover a short time-scale of less than $1{\rm s}$, a plausible mechanism is needed to dampen the enhanced magnetic field. 

It is worth noting that, the prior of $M_{\rm TOV}$ in this work is directly extracted from \cite{miller2021} and \cite{li2021}. \cite{Raaijmakers2021} independently analized the NICER data and found a quite different probability density distribution for $M_{\rm TOV}$ on compared with \cite{miller2021}. They claimed that $M_{\rm TOV}$ is smaller that $2.4M_{\odot}$ at $95\%$ confidence level. Combined with the mass measurement of J0952-0670 ($\sim 2.35M_{\odot}$), $M_{\rm TOV}$ should be well constrained near $2.3M_{\odot} - 2.4M_{\odot}$. A long-lived remnant is unlikely to be produced after GW170817 and the GW-FRB association is not real.


\section*{Acknowledgements}
This work is supported by the China Postdoctoral Science Foundation (2023M732713), the National Natural Science Foundation of China (Projects 12021003), and the National SKA Program of China (2022SKA0130100).

\section*{Data Availability}
The BAT and XRT light curve data for sGRBs are obtained by using the public data from the Swift archive: \url{https://www.swift.ac.uk/burst\_analyser/}. The masses for GW170817 and GW190425 are obtained by using the the public data from LIGO-Virgo-Kagra Collaboration through:
\url{https://dcc.ligo.org/LIGO-P1800370/public} and \url{https://dcc.ligo.org/LIGO-P2000026/public}.







\appendix
\section{internal X-ray plateaus}
\label{sec:fx}
\begin{figure}
\resizebox{80mm}{!}{\includegraphics{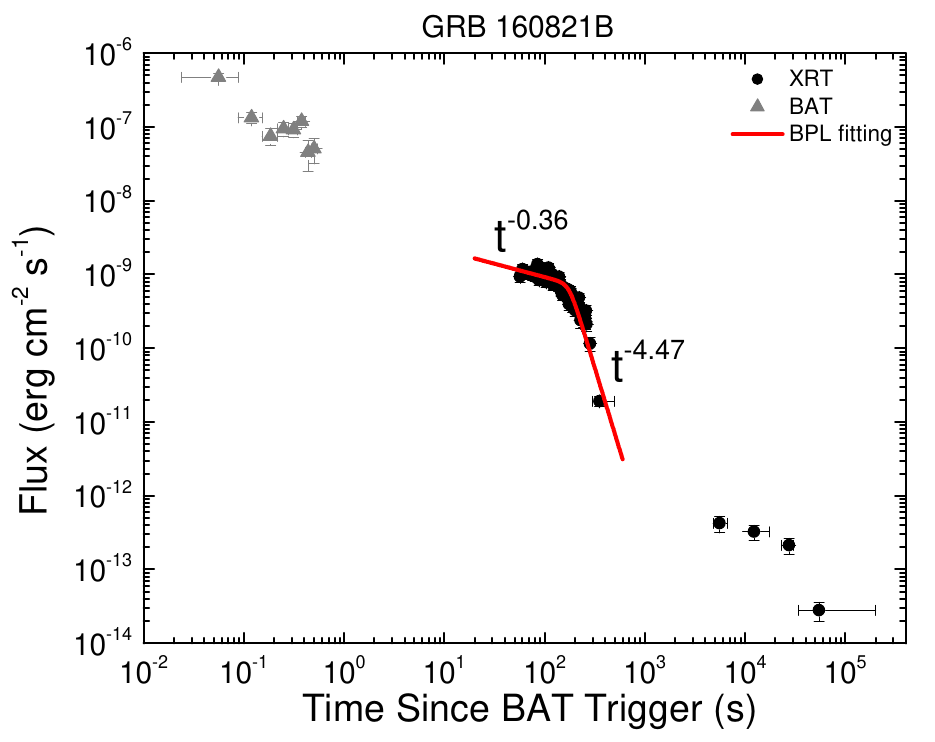}}
\caption{An example of internal X-ray plateau light-curve fitting for GRB 160821B. The red solid line shows the broken power-law function fits, and the black circles and gray triangles represent XRT and BAT observation data, respectively.}
\label{fig:internal_plateau sample}
\end{figure}

We analyzed 144 sGRBs observed with Swift from 2005 January to 2023 January. In principle, we geometrically extrapolated the BAT-band (15--150 keV) data to the XRT-band (0.3--10 keV) by assuming a single power-law spectrum, and then performed a temporal fit to the combined light curve with a smooth broken power-law function to search for a possible plateau feature (shallow decay slope smaller than 0.5). In particular, we focused on collecting those sGRBs that exhibit a plateau followed by a decay index steeper than 3 as our internal plateau sample. The steeply decay slope would suggest a sudden cessation of central engine activity, which is very likely due to the collapse of a supra-massive NS into a BH. We finally found 35 sGRBs with internal plateau characteristic, which comprises 24\% in the entire short GRBs' sample. An example of the internal X-ray plateau light-curve fitting is shown in Figure \ref{fig:internal_plateau sample}. Remarkably, this fraction should be considered as a lower limit for supra-massive NSs formation fraction from binary neutron star mergers, since some sGRBs may come from NS + BH mergers or some supermassive NSs collapsed so late that the relevant data were missed.

\section{Mass distribution of Galactic DNS systems}
\label{sec:BNS_mass}
\begin{figure}
    \resizebox{80mm}{!}{\includegraphics{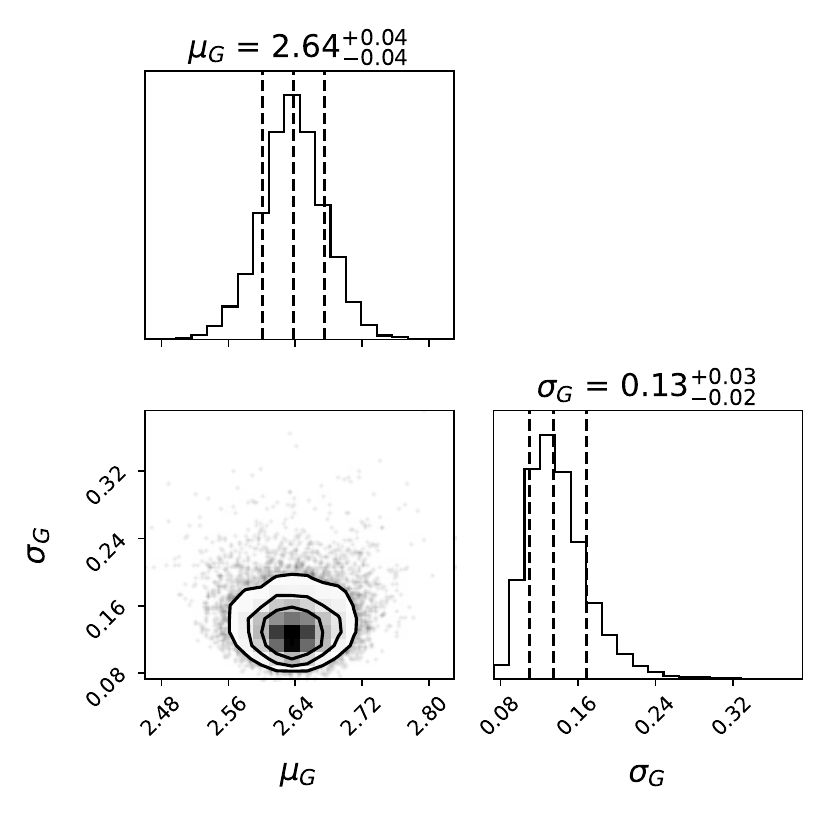}}
    \caption{The probability density distribution of parameters to describe the mass distribution of Galactic DNSs. The titles show the central value and uncertainties at $1\sigma$ confidence level.}
    \label{fig:mass_distribution_parameter}
\end{figure}

We collected the confirmed and candidate Galactic DNS (containing at least one pulsar) systems with total mass measured in Table \ref{tab:BNS_list}. Then we use a Gaussian function with mean value $\mu_G$ and standard deviation $\sigma_G$ to characterize the distribution of their total masses. Using the Bayesian inference MCMC simulation, the probability density distribution of $\mu_G$ and $\sigma_G$ can be obtained, which are shown in Figure \ref{fig:mass_distribution_parameter}.

\begin{table}
\caption{List of confirmed and candidate Galactic DNS systems.}
\centering
\begin{tabular}{|c|c|c|}
  \hline
Pulsar & $M_{\rm tot}~(M_{\odot})$  & Reference   \\\hline
J0453+1559 & 2.734(4) & \makecell[c]{\cite{deneva2013}\\ \cite{martinez2015}}  \\ \hline
J0509+3801 & 2.805(3) & \cite{Lynch2018} \\ \hline
J0514-4002A & 2.4730(6) & \makecell{\cite{freire2007} \\ \cite{ridolfi2019}}  \\ \hline
J0737-3039A & \multirow{2}*{2.587052(+9/-7)} & \multirow{2}*{\cite{kramer2006}} \\ 
J0737-3039B &  &\\ \hline
J1325-6253 & 2.57(6) & \cite{sengar2022}\\ \hline
J1411+2551 & 2.538(22) & \cite{martinez2017}\\ \hline
J1518+4904 & 2.7183(7) & \cite{janssen2008} \\ \hline
B1534+12 & 2.678463(4) & \cite{fonseca2014} \\ \hline 
J1748-2021B & 2.92(15) & \cite{freire2008} \\ \hline
J1756-2251 & 2.56999(6) & \cite{ferdman2014}\\ \hline
J1757-1854 & 2.732882(12) & \cite{cameron2018,cameron2023}\\ \hline
J1759+5036 & 2.62(3) & \cite{agazie2021} \\ \hline
J1807-2500B & 2.57190(73) & \cite{lynch2012} \\ \hline
J1811-1736 & 2.57(10) & \cite{corongiu2007}\\ \hline
J1823-3021G & 2.65(7) & \cite{ridolfi2021}\\ \hline
J1829+2456 & 2.60551(38) & \cite{champion2005} \\ \hline
J1906+0746 & 2.6134(3) & \cite{van_Leeuwen2015}\\ \hline
J1913+1102 & 2.8887(6) & \cite{ferdman2020}\\ \hline
B1913+16 & 2.828378(7) & \cite{weisberg2010} \\ \hline
J1930-1852 & 2.54(3) & \cite{swiggum2015} \\ \hline
J1946+2052 & 2.50(4) & \cite{stovall2018}\\ \hline
B2127+11C & 2.71279(13) & \cite{jacoby2006} \\ \hline
\end{tabular}
\label{tab:BNS_list}
\end{table}

\section{total masses and mass ratios of GW170817 and GW190425}
\label{sec:Mtot_q}
\begin{figure*}
    \centering
    \begin{tabular}{ll}
    \resizebox{80mm}{!}{\includegraphics{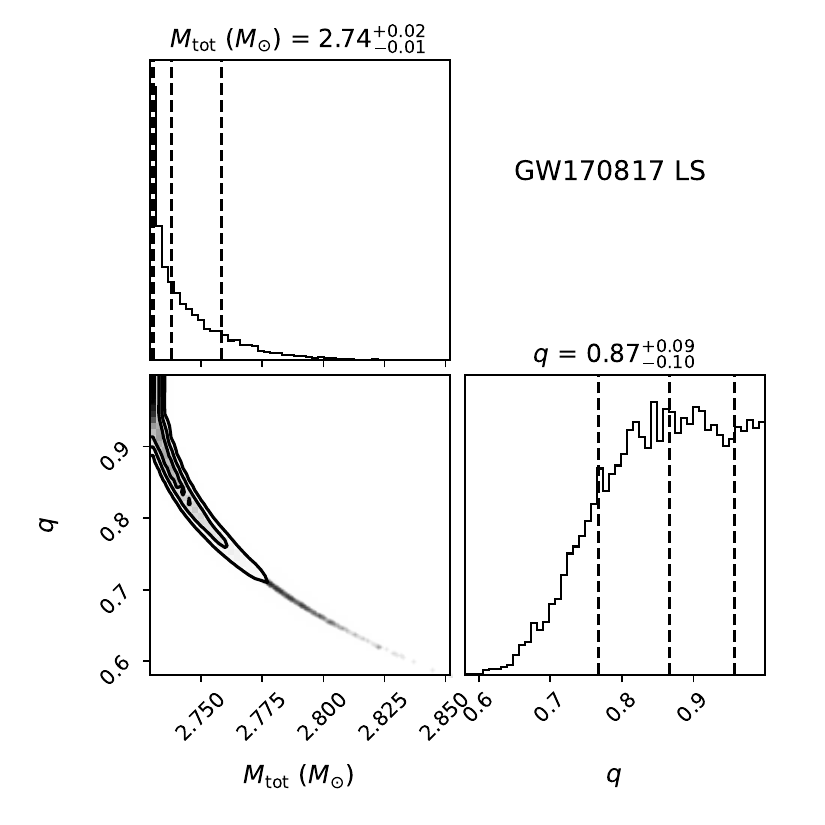}} &
    \resizebox{80mm}{!}{\includegraphics{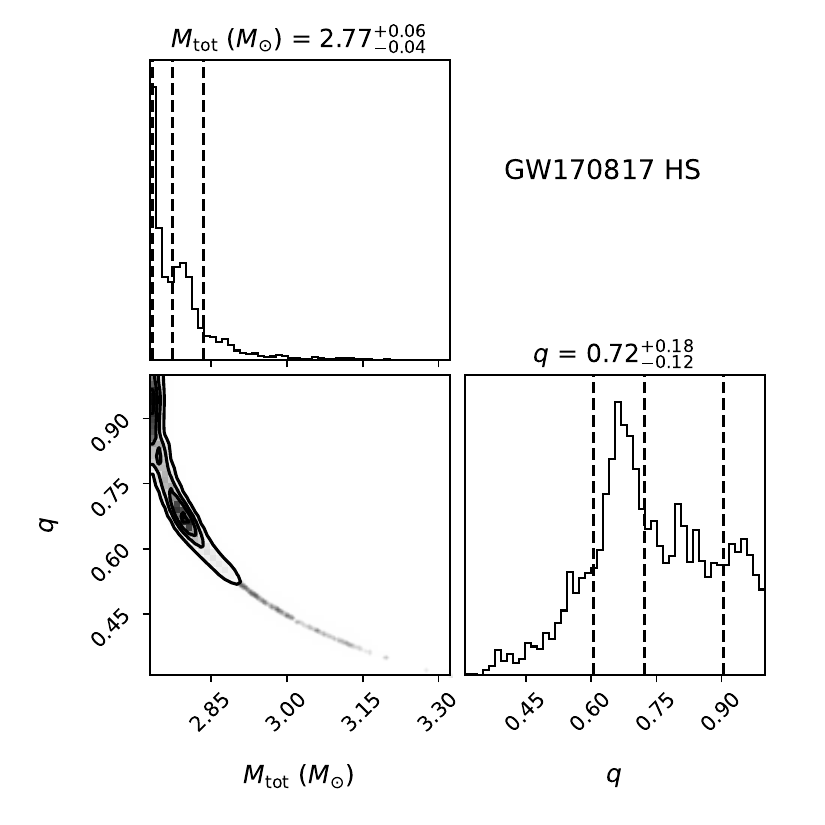}}
    \\
    \resizebox{80mm}{!}{\includegraphics{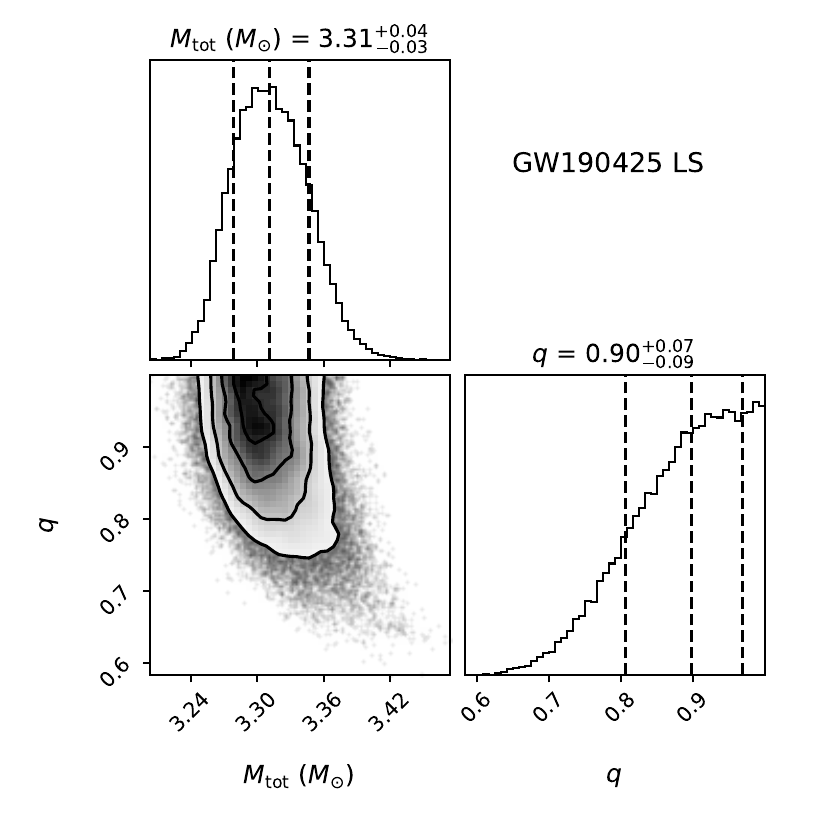}} &
    \resizebox{80mm}{!}{\includegraphics{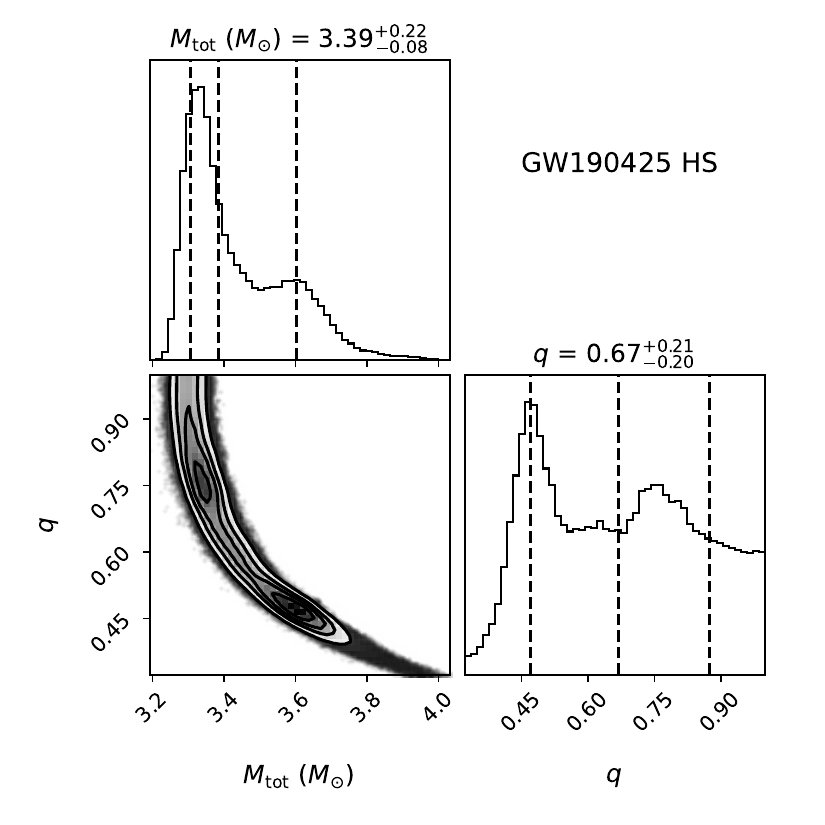}}
    \end{tabular}
    \caption{The joint probability density distributions of the total masses and mass ratios of the BNS-merger systems GW170817 and GW190425 in the source frame.}
    \label{fig:mass_q}
\end{figure*}

The posteriors of the parameter estimations for GW170817 and GW190425. For GW170817, the posteriors of $M_{d,1}$ and $M_{d,2}$ in the detector's frame. Considering that the redshift of GW170817 is known ($z = 0.008$) \citep{abbott2017b}, the masses in the source frame are be calculated as $M_{i} = M_{d,i} / (1 + z)$. For GW190425, the masses in the source frame are directly given. The values shown in Figure \ref{fig:mass_q} are all defined in the source frame. 

\section{Parameter spaces of Equation of states used to construct universal relations}
\label{sec:parameter_space}
\begin{figure}
	\includegraphics[width=\columnwidth]{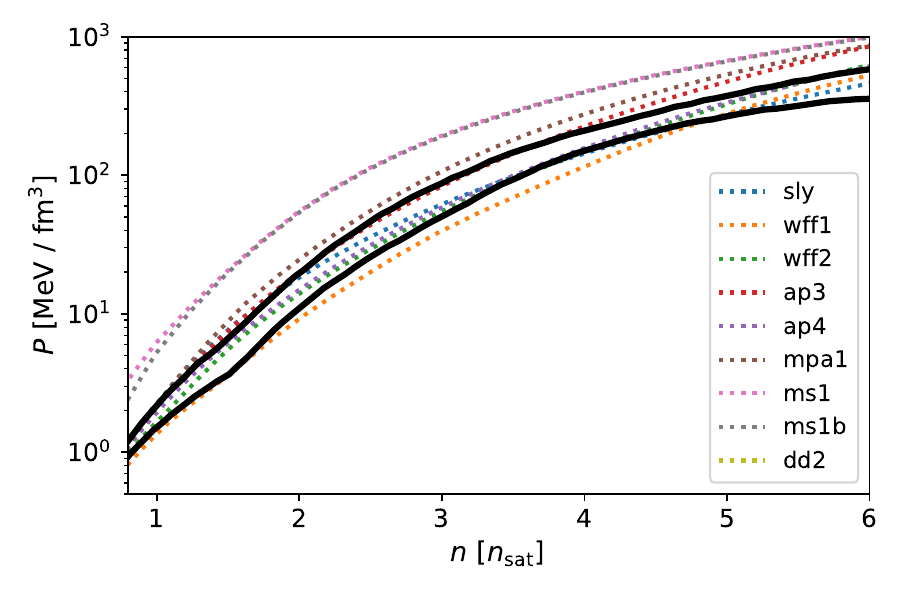} \\
    \includegraphics[width=\columnwidth]{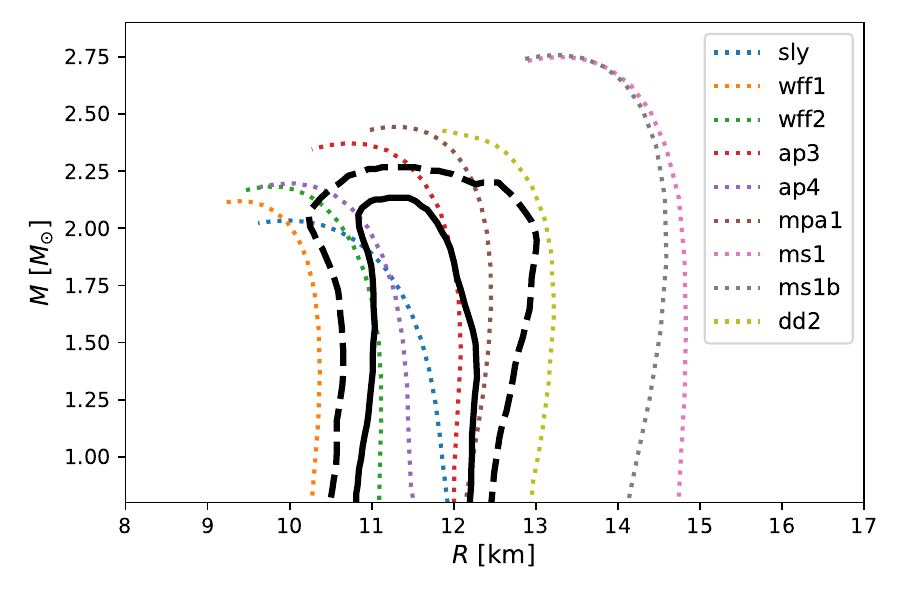}
    \caption{The microcosmic and macroscopic parameter space for NS EoSs. The black solid and dashed lines are extracted from \citet{miller2021}, which show the allowed parameter spaces at $68\%$ and $95\%$ confidence levels, respectively. The colored dotted lines show the $P$-$n$ and $M$-$R$ relations for the EoSs used to build the universal relation for $M$-$M_b$ conversion in \citet{gao2020}.}
    \label{fig:EOS_parameter_space}
\end{figure}

\cite{gao2020} used a collection of common NS EoSs in the literature to build the universal relations of the gravitational and baryonic masses of non-rotating and rapidly rotating NSs. Their $P$-$n$ and $M$-$R$ relations are shown in Figure \ref{fig:EOS_parameter_space}, together with the allowed parameter spaces for NS EoS constrained in \cite{miller2021}. Please find the list the EoS names, $M_{\rm TOV}$ and $R_{1.4}$ values in \cite{gao2020} and the references therein. Generally, the EoSs used in \cite{gao2020} can well cover the allowed parameter space in \cite{miller2021}, so that the use of their proposed universal relations is conservative.

\bsp	
\label{lastpage}
\end{document}